\documentclass[journal]{IEEEtran}

\usepackage{graphicx}
\usepackage{booktabs}
\usepackage{bbding}
\usepackage{subfigure}
\usepackage{overpic}
\usepackage{amsmath}
\usepackage{amssymb}
\usepackage{multirow}
\usepackage{makecell}
\usepackage{array}
\usepackage{stfloats}
\usepackage{algorithm,amsfonts}
\usepackage{subfigure}
\usepackage{algpseudocode}
\usepackage{amsmath,epsfig, hyperref}
\usepackage{tabu}

\usepackage{soul}
\soulregister{\ref}7

\usepackage{makecell}

\hypersetup{
	colorlinks=true,
	linkcolor=cyan,
	urlcolor=red,
	citecolor=cyan,
}

\begin{document}

\title{Change Detection from Synthetic Aperture Radar Images via Dual Path Denoising Network}
\author{Junjie Wang, Feng Gao, Junyu Dong, Qian Du, Heng-Chao Li

\thanks{This work was supported in part by the National Key Research and Development Program of China under Grant 2018AAA0100602, in part by the Key Research and Development Program of Shandong Province under Grant 2019GHY112048, and in part by the Natural Science Foundation of Shandong Province under Grant ZR2019QD011. (\emph{Corresponding author: Feng Gao})}

\thanks{
Junjie Wang, Feng Gao, and Junyu Dong are with the School of Computer Science and Technology, Ocean University of China, Qingdao 266100, China.

Qian Du is with the Department of Electrical and Computer Engineering, Mississippi State University, Starkville, MS 39762 USA.

Heng-Chao Li is with the School of Information Science and Technology, Southwest Jiaotong University, Chengdu 611756, China.}
}

\markboth{}%
{Shell}

\maketitle
\begin{abstract}

Benefited from the rapid and sustainable development of synthetic aperture radar (SAR) sensors, change detection from SAR images has received increasing attentions over the past few years. Existing unsupervised deep learning-based methods have made great efforts to exploit robust feature representations, but they consume much time to optimize parameters. Besides, these methods use clustering to obtain pseudo-labels for training, and the pseudo-labeled samples often involve errors, which can be considered as ``label noise". To address these issues, we propose a \underline{D}ual \underline{P}ath \underline{D}enoising \underline{Net}work (DPDNet) for SAR image change detection. In particular, we introduce the random label propagation to clean the label noise involved in preclassification. We also propose the distinctive patch convolution for feature representation learning to reduce the time consumption. Specifically, the attention mechanism is used to select distinctive pixels in the feature maps, and patches around these pixels are selected as convolution kernels. Consequently, the DPDNet does not require a great number of training samples for parameter optimization, and its computational efficiency is greatly enhanced. Extensive experiments have been conducted on five SAR datasets to verify the proposed DPDNet. The experimental results demonstrate that our method outperforms several state-of-the-art methods in change detection results.

\end{abstract}


\IEEEpeerreviewmaketitle

\section{Introduction}

\IEEEPARstart{O}{wing} to the sustainable development of remote sensing imaging science and technology, plenty of synthetic aperture radar (SAR) sensors are designed for aerial and spaceborne platforms. The ability to collect images of the same locations at higher frequency allows near real-time monitoring of the Earth, and therefore increasingly SAR images of the same geographical area are available. These images provide insights to many applications, such as change detection \cite{Radke05} \cite{campos20grsl} \cite{vinholi20grsl} \cite{lv18tii}, natural disaster monitoring \cite{zhuang21tgrs} \cite{dong21tgrs} \cite{Brunner10}, urban planning \cite{Quan18} \cite{wang21jstars} \cite{zou21grsl}, and land cover data updating \cite{Zanotta12}. Among these applications, change detection aims to recognize the changed information of the same region by analyzing mutlitemporal images. It has attracted widespread interest in recent years \cite{amitrano21tgrs}.

\begin{figure}[ht]
\begin{center}
\includegraphics [width=2.8in]{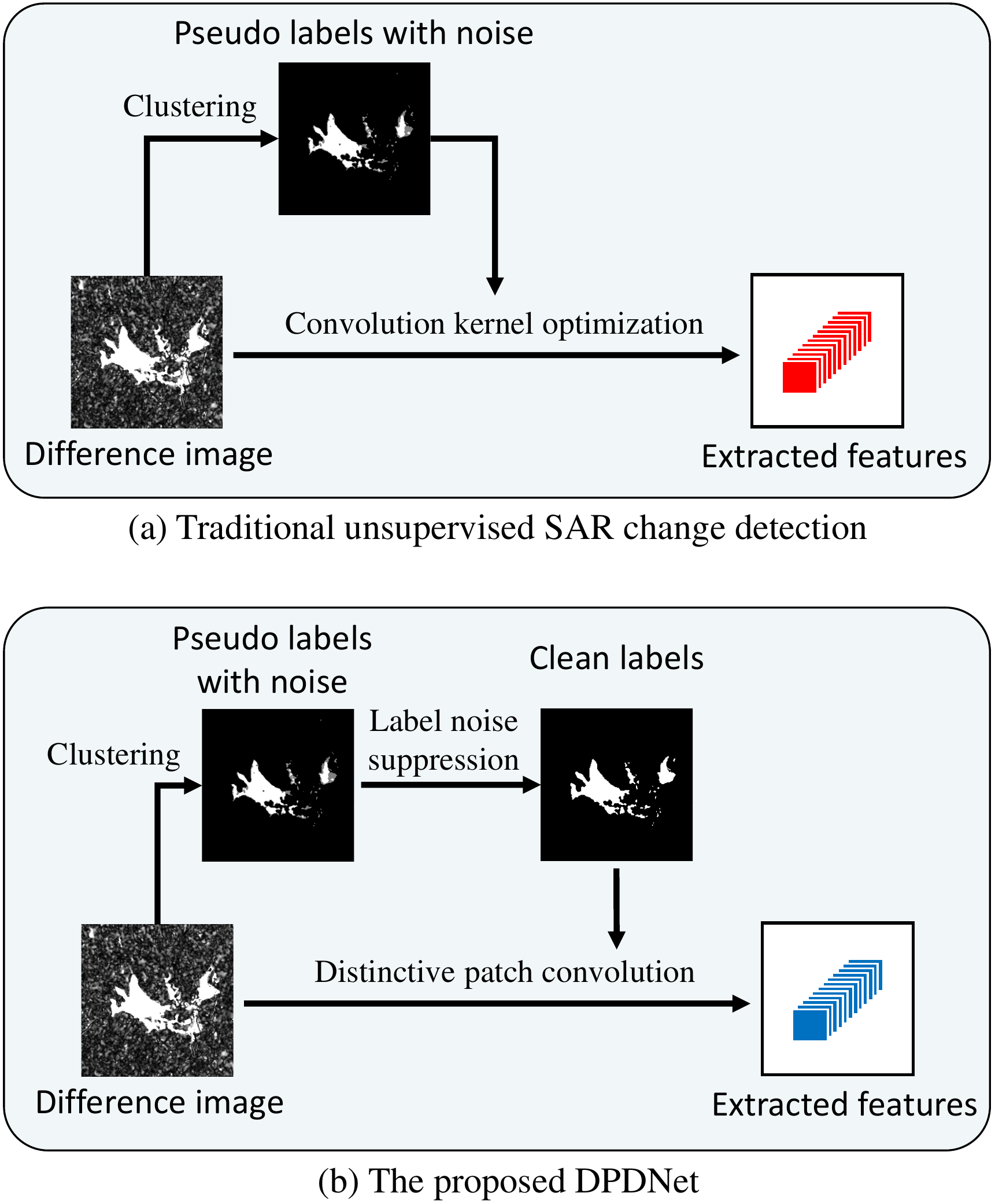}
\caption{Comparisons of traditional methods with the proposed DPDNet. (a) The general framework of existing unsupervised SAR image change detection method. These methods contain many noise labels and are commonly time-consuming in the training phase. (b) The proposed DPDNet. It generates clean labels and is more efficient in the training phase.}
\label{modelcompare}
\end{center}
\end{figure}

SAR change detection is commonly considered as a classification task. In a general pipeline of SAR change detection, the difference image (DI) is first created, and the DI is classified into changed or unchanged pixels in a supervised or unsupervised manner \cite{sumaiya16grsl}. However, the existence of speckle noise is a non-negligible problem in SAR change detection task. Speckle noise in SAR images appears as a form of multiplicative noise and degrades the image quality. False alarms are usually produced due to the existence of speckle noise \cite{yang19tgrs} \cite{wu19multitemp}. Therefore, the DI classification for SAR images is still a challenging task \cite{li19grsl} \cite{pham16tgrs}. So, it is critical to design robust change detection techniques which are effective in speckle noise suppression \cite{saha21tgrs} \cite{marin15tgrs} \cite{an15grsl}.

In the past few years, researchers have devoted great efforts towards solving or alleviating the effect of speckle noise. There are two types of change detection techniques: supervised and unsupervised methods. Supervised methods require labeled training samples to learn the parameters of a classifier \cite{Volpi13} \cite{Wang16}. Compared with it, unsupervised methods are more popular since they compare two multitemporal SAR images without any prior knowledge or manually labeled samples \cite{Bruzzone00} \cite{Yetgin12} \cite{Gong17isprs}. In this paper, we mainly focus on developing unsupervised SAR image change detection method.

Existing unsupervised change detection methods mainly focus on DI generation and classification. In DI generation, log-ratio \cite{bovolo05tgrs}, Gauss-ratio \cite{Hou14} or neighborhood-based ratio \cite{Gong12_grsl} operators are used. Besides, the coefficient of variation based on time series SAR images is used to avoid the speckle noise \cite{Jiangisprs20}. For DI classification,  many clustering methods have been employed, such as fuzzy $c$-means (FCM) \cite{Mishra12}, $k$-means clustering \cite{Celik09}, multiple kernel clustering \cite{Jia16}, and mean-shift algorithms \cite{Aiazzi13}. In addition, the machine learning methods based on Markov random fields are also applied to the despeckle task \cite{Bruzzone00}.

To further enhance the DI classification performance, researchers incorporate deep learning-based classifiers into the traditional unsupervised model. Gong et al. \cite{Gong16_tnnls} first assigned pseudo-labels to the pixels in DI by an FCM-based joint classifier. Then restricted Boltzmann machines (RBMs) were trained for change map generation. Hou et al. \cite{hou17grsl} presented a change detection method by combining saliency computation and low-rank algorithm. Zhan et al. \cite{Zhan17_grsl} proposed a deep Siamese CNN model to extract discriminant features for change detection. Zhang et al. \cite{zhang21grsl} presented a deep spatial-temporal gray-level co-occurrence CNN, which is capable of exploiting the spatial-temporal information of mutlitemporal images. Dong et al. \cite{dong21tgrs} integrated unsupervised clustering with CNN to learn clustering-friendly feature representations from multitemporal SAR images. Qu et al. \cite{qu21grsl} proposed a dual-domain network. Features from both frequency and spatial domains are exploited to alleviate the speckle noise.

As illustrated in Fig. \ref{modelcompare} (a), existing deep learning-based unsupervised SAR change detection methods confront two challenges: 1) \textbf{The obtained pseudo-labels commonly involve errors.} The existing methods generally use clustering to obtain pseudo-labels, which contain some errors. This phenomenon is called ``label noise", which can affect the subsequent network optimization. Accordingly, it is highly desired to devise a label noise tolerant method. 2) \textbf{Deep learning-based methods are time-consuming in the training phase.} To obtain satisfying results, the existing methods run many epochs during the training phase or utilize the pretraining strategy to obtain the desired features (Here, the pretraining time is also regarded as a part of time consumption). Both of them are time-consuming and require reliable training samples, which are quite limited in SAR change detection. Therefore, how to design a network that is less time-consuming and requires fewer training samples has become the key to the research.

To tackle the above-mentioned problems, we propose a \underline{D}ual \underline{P}ath \underline{D}enoising \underline{Net}work (DPDNet) to alleviate the effect of speckle noise and label noise simultaneously (see Fig. \ref{modelcompare} (b)). As illustrated in Fig. \ref{structure}, one branch of DPDNet uses random label propagation to clean the label noise. Besides, the other branch of DPDNet is employed to extract shallow and deep features, and then combines them together for hierarchical feature representation. In this process, the designed distinctive patch convolution (DPConv) is utilized to simplify the network structure and reduce the training time. To validate the effectiveness of the proposed DPDNet, we conducted extensive experiments on five datasets. Moreover, we compared our method with six state-of-the-art models, and the experimental results demonstrate that our method can reduce the effect of speckle noise and label noise effectively. Besides, the DPDNet consumes less time than other models while achieving satisfying change detection results.

In summary, the main contributions of this paper are as follows:

\begin{enumerate}

  \item To the best of our knowledge, we make the first attempt to solve the problem of speckle noise and label noise simultaneously. Previous works mainly focus on the speckle noise. However, the label noise problem is detrimental to the change detection performance. The proposed DPDNet can alleviate two kinds of noise simultaneously, and generate more accurate change maps.

  \item We present a novel distinctive patch convolution (DPConv), which simplifies the network structure and expedite the training phase. The patches taken from the original images are used as convolutional kernels, and therefore it does not require many training samples for parameter optimization.

\end{enumerate}

\section{Methodology}

\begin{figure*}
\centering
\begin{center}
\includegraphics [width=6.5in]{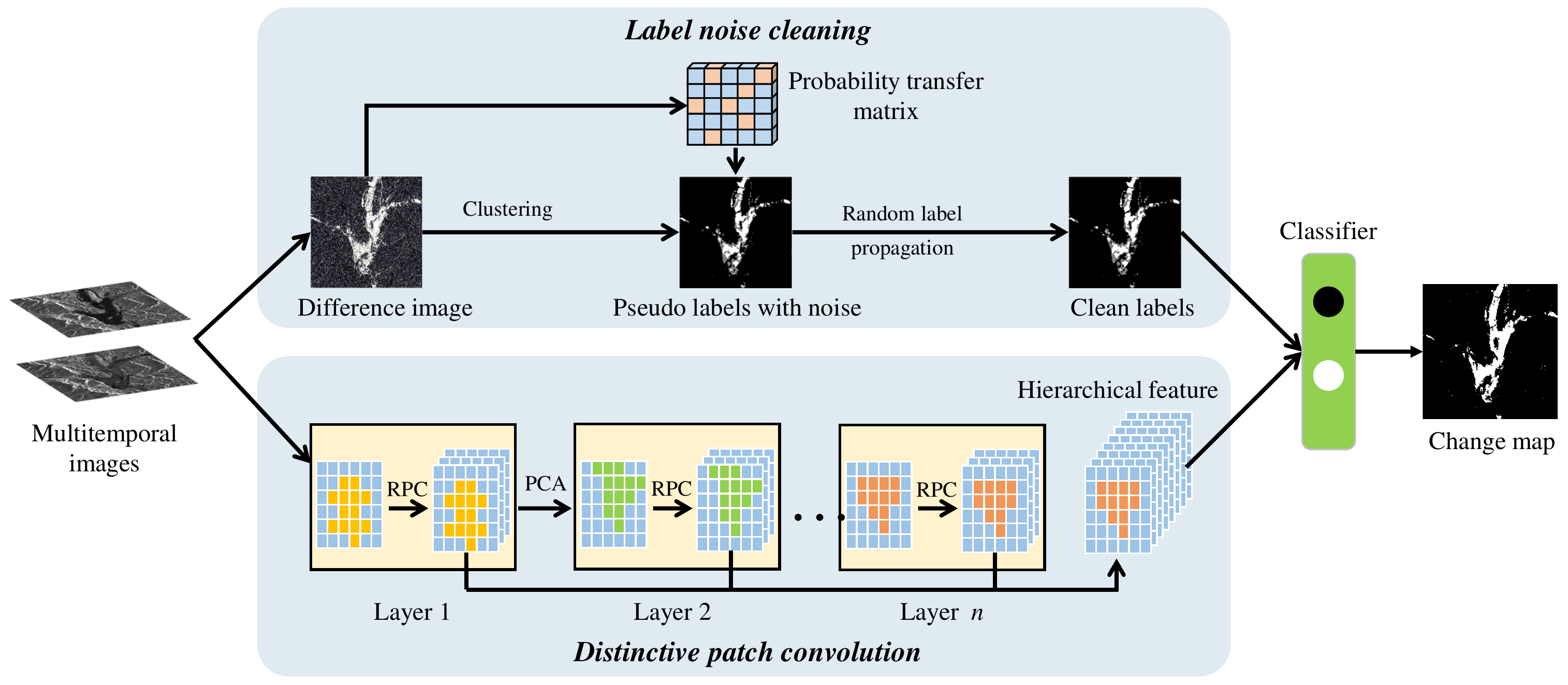}
\caption{Schematic illustration of the proposed DPDNet. One branch uses random label propagation to clean the label noise generated in preclassification. The other branch is utilized to extract shallow and deep features by stacking multiple convolution layers. Finally, the clean labels and stacked features are fed into the classifier to compute the final change detection results.}
\label{structure}
\end{center}
\end{figure*}

As shown in Fig. \ref{structure}, the proposed DPDNet comprises two branches: one is used to clean the label noise by propagating the correct labeled samples to the other unlabeled samples. In this process, the transformation weight is generated through homogeneous region generation and similarity measurement, followed by multiple random label propagation based on the weights and using voting to obtain the final labels. The other branch aims to improve the efficiency of the model while avoiding the speckle effect. To reduce the long training time required by conventional convolution, the designed DPConv is used to obtain hierarchical convolution features, so as to reduce the consumption of computational resources while improving feature expressiveness. Through the combination of these two branches, it not only enables the removal of noise from the data and improves the change detection performance, but also reduces the training time and improves the application potential. 

\subsection{Overview of the Proposed DPDNet}

\begin{figure}[ht]
\begin{center}
\includegraphics [width=2.8in]{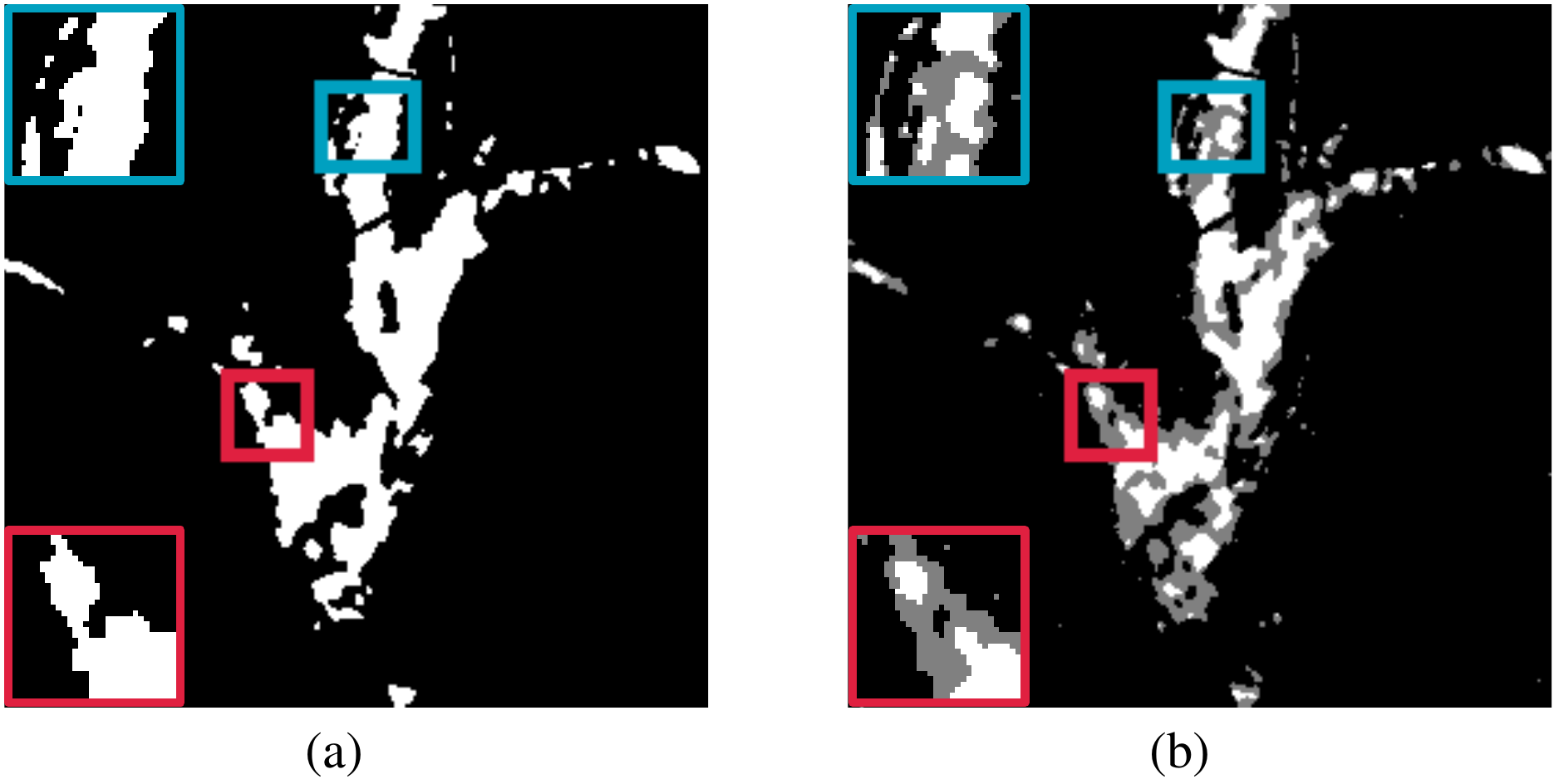}
\caption{Visualization of label noise. (a) The ground-truth change map. (b) The pseudo-labeled change map generated by $k$-means \cite{Celik09} clustering. From the marked areas, it can be observed that there some differences between the ground-truth map with the generated pseudo-labeled map. If these noisy labels are fed into the neural network, they will mislead the training process, and cause a decrease in the change detection performance.}
\label{cluster_noise}
\end{center}
\end{figure}

Most existing deep learning-based methods applied to SAR change detection tasks require multiple rounds of training for parameter optimization, which is very time-consuming. Besides, in the process of clustering, considerable noisy samples are inevitably introduced due to the limitation of clustering algorithm (see Fig. \ref{cluster_noise}), which is rarely considered in previous works \cite{qu21grsl} \cite{Gao21capsule} \cite{Gong17isprs}. However, a robust change detection method should take label noise into account. To solve the above two problems, a dual path change detection model has been proposed, which is shown in Fig. \ref{structure}. The first branch is used to guide the removal of label noise. Its core idea is to randomly remove the labels of some selected samples and repredict the labels of these selected samples based on the prior knowledge and spatial correlation. The other branch focuses on despeckle, and the proposed DPConv is used replacing the conventional convolution operation to reduce the time consumption while extracting meaningful features. The experimental results show that the combination of the two branches is an effective solution to the above problems and has a certain improvement both in terms of efficiency and evaluation indicators compared to other methods.

\subsection{Label Noise Cleaning with Random Label Propagation}

\begin{figure}[ht]
\begin{center}
\includegraphics [width=3.5in]{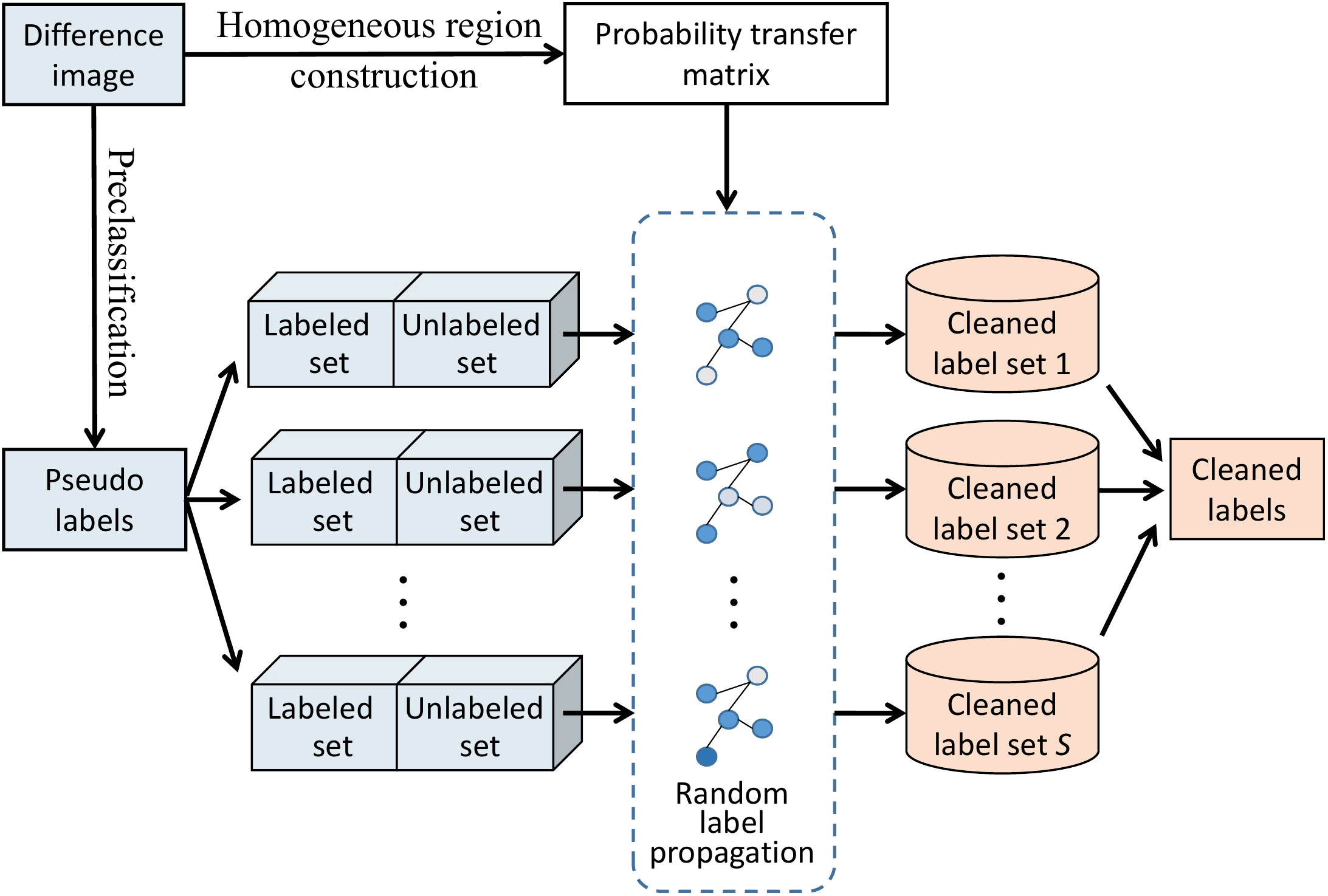}
\caption{Schematic of the proposed RLPA-based label noise cleaning for SAR change detection.}
\label{noisy_label}
\end{center}
\end{figure}

As illustrated in Fig. \ref{cluster_noise}, there are some differences between the ground-truth change map with the generated pseudo-labeled change map. These differences can be considered as the label noise. If these noisy labels are fed into the neural network, they will mislead the training process and thus cause a decrease in the final change detection performance. To effectively clean the label noise, we introduce the random label propagation algorithm (RLPA) \cite{Jiang19} in our network, which is shown in Fig. \ref{noisy_label}. It mainly includes three phases: homogeneous region construction, generation of probabilistic transition matrix, and random label propagation.

\emph{Homogeneous Region Construction:} We aim to propagate the label information among samples to clean the label noise. We assume that pixels from the same homogeneous region are more likely to belong to the same class. Thus, the construction of homogeneous regions plays an important role in label noise cleaning. The entropy rate superpixel segmentation \cite{liu11cvpr}, which is an objective function of superpixel segmentation, is employed to construct homogeneous regions. Through this function, the whole image can be divided into several regions, and each region is recognized as a homogeneous region to facilitate subsequent label propagation.

\emph{Probabilistic Transition Matrix Generation \cite{Xu18}:} Upon the acquired homogeneous regions, we can assign a weight to the pixels within a homogeneous region based on the spatial similarity, and let the weight between the pixels from different homogeneous regions be zero.

\begin{equation}
\begin{aligned}
{W_{ij}} = \left\{ \begin{array}{ccccc}
\exp ( - \frac{{\textrm{sim}{{({p_i},{p_j})}^2}}}{{2{\sigma ^2}}}),{\kern 1pt} {\kern 1pt} {\kern 1pt} {\kern 1pt} {\kern 1pt} {\kern 1pt} {\kern 1pt}  & {p_i},{p_j} \in {\chi _l}\\
0, & {p_i} \in {\chi _l}{\kern 1pt} {\kern 1pt} \textrm{and} {\kern 1pt} {\kern 1pt} {p_j} \in {\chi _p},
\end{array} \right.
\end{aligned}
\end{equation}
where $sim({p_i},{p_j})$ represents the spatial similarity, $\chi _l$ and $\chi _p$ are the $l$th and $p$th homogeneous region, and $\sigma$ is the mean variance of all pixels in each homogeneous region.

In this paper, the Euclidean distance is used to measure the similarity

\begin{equation}
\textrm{sim}({p_i},{p_j})=\lVert p_i-p_j\rVert_2,
\end{equation}
where $\lVert\cdot\rVert_{2}$ demotes the $l_2$ norm.

After that, the label information can propagate through the connected weight between different pixels. The larger the weight between two nodes, the easier it is likely to traverse. Hence, the two nodes are more likely to belong to the same class. We can define a probabilistic transition matrix $T$ which can be seen as the probability of label conversion.

\begin{equation}
T_{ij}=P(j \to i)=\frac{W_{ij}}{\sum_{k=1}^{N} W_{kj}},
\end{equation}
where $W_{ij}$ is the connected weight between different pixels. Here, $N$ is the number of pixels in the homogeneous region, and $k$ traverses all pixels in the homogeneous region, and $T_{ij}$ can be considered as the probability to travel from node $j$ to $i$.

\emph{Random Label Propagation:} The training dataset $X$ is divided into a labeled subset
$X_L =\{ x_1,x_2,...,x_l\} \in R^l$ ($X_L$ is the set of real vector of $l$ elements) and an unlabeled subset
$X_U = \{ x_{l+1},x_{l+2},...,x_N\} \in R^{N-l}$ ($X_U$ denotes the set of unlabeled vector of $N-l$ elements). The label matrix of $X_L$ is denoted as $\hat{Y}_L$, while the labels of $X_U$ are discarded. In the experiments, we randomly choose two subsets from the total training samples. Afterwards, the task turns to be predicting the labels $\hat{Y}_U$ of the unlabeled subset $X_U$ based on the generated probabilistic transition matrix. To better propagate the label information, multiple iterations are adopted to propagate the labels from $X_L$ to $X_U$. Hence, the label of $x_i$ at time $t+1$ is:

\begin{equation}
y_i^{t+1}=\alpha \sum_{x_i,x_j\in X} T_{ij}y_j^t+(1-\alpha)\widetilde{y_i}^{LU},
\end{equation}
where $\alpha$ is a parameter to balance the contribution between the initial label information and the label information received from its neighbors, and $\widetilde{y_i}^{LU}$ is the $i^\textrm{th}$ column of $\widetilde Y_{LU} = [\widetilde Y_L \widetilde Y_U]^T$. It should be noted that the labels of the unlabeled subset are initialized to zero.

Since the initial labeled and unlabeled subsets are chosen randomly, the above-mentioned process of random assignment of clean samples and unlabeled samples are repeated multi times. Because each sample gets one label per round, each sample will obtain multiple labels after multiple rounds of random assignment. The final label can be computed by majority vote.

\subsection{Alleviating Speckle Noise with Distinctive Patch Convolution}

Speckle noise is a granular interference that inherently degrades the quality of SAR images, and it severely affects the change detection performance. Many deep learning-based classifiers have been used to suppress the speckle noise \cite{shen21tgrs} \cite{ma20tgrs}. However, these techniques are commonly time-consuming in the training period. Besides, these techniques generally require a great number of training samples for model training to achieve robust representation. To solve the mentioned issue, we propose a distinctive patch convolution (DPConv) for SAR image feature representation.

Xu et al. \cite{Xu18} proposed a random patch network (RPNet), which utilizes the random patches taken from the image to build a random matrix, and then project the original vector into a proper subspace. This work reveals the potential value of extracting patches from the original images as convolution kernels, which has not been considered in SAR image change detection. However, there are certain problems with this approach, despite the advances it achieved. In \cite{ouyang19}, it points out that in a structured model, the randomness of sampling can risk the model's capability of focusing on non-trivial spatial contextual information. As we all know, SAR image is often composed of background and semantic objects. The texture and geometry information of semantic objects is distinctive, so these semantic objects (called distinctive region) commonly have higher activation values in the feature map. Therefore, when the convolutional kernels are selected from the background, they do not contain as much information as the semantic object regions, and this view is also demonstrated by comparative experiments in Section~\ref{sec1}. To solve the problem, we proposed DPConv, which selects patches from the distinctive regions as convolutional kernels to avoid the randomness of convolution kernel selection, so as to focus on important regions with salient texture and geometry information.

\begin{figure}[ht]
\begin{center}
\includegraphics [width=3.3in]{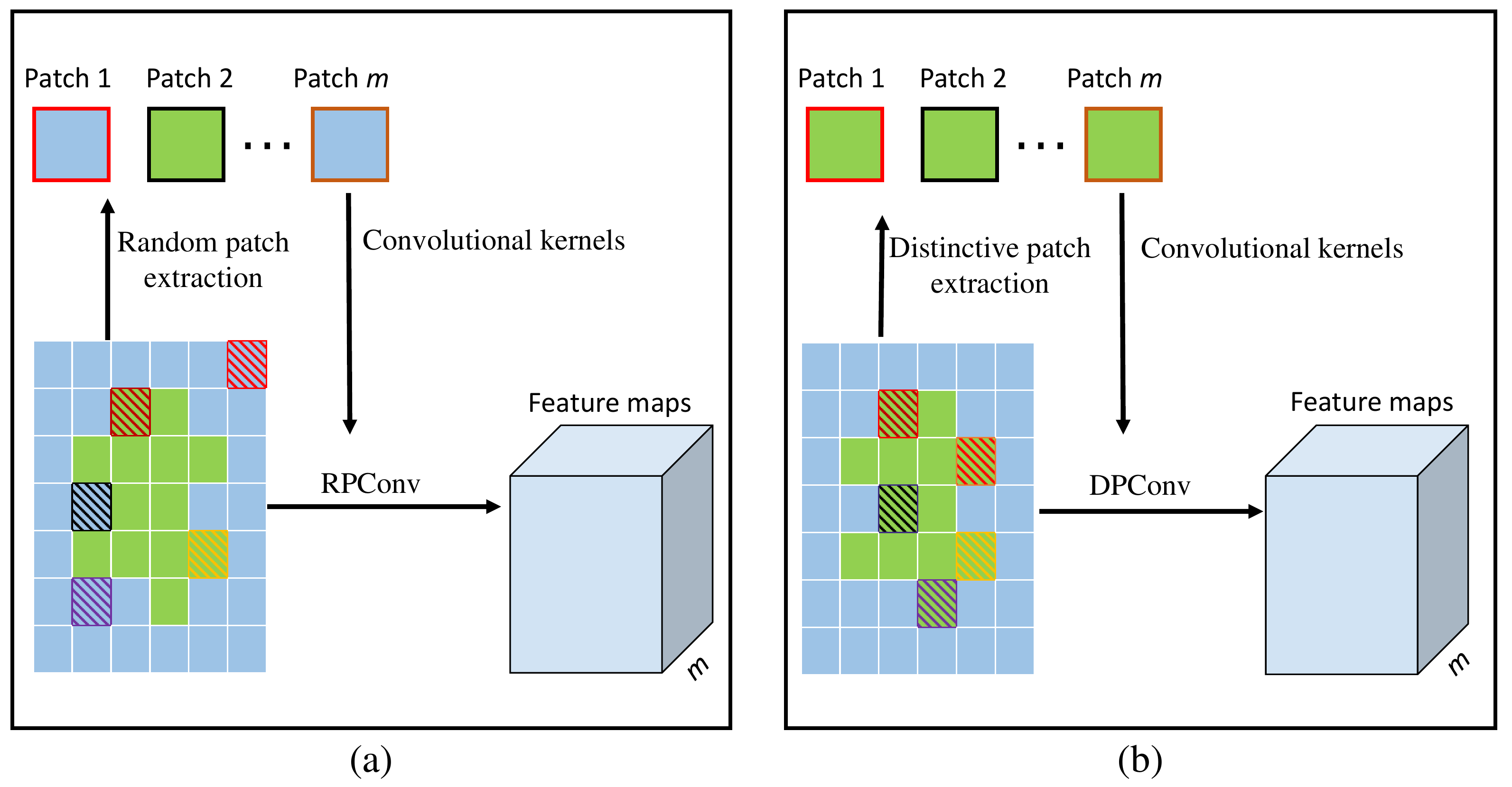}
\caption{Comparisons of RPConv with the proposed DPConv. (a) represents the feature extraction by RPConv. (b) represents the feature extraction by DPConv. The salient object is marked by green squares, the background is marked by blue square. The selected regions are marked by diagonal squares. By using RPConv, background region which contains little texture are chosen. In contrast, by using the proposed DPConv, convolutional kernels are selected from the distinctive region, and therefore boosted classification performance can be achieved.}
\label{layer}
\end{center}
\end{figure}

The implementation details of DPConv are shown in Fig. \ref{layer}. The distinctive regions are obtained according to the activation values in the feature maps as:
\begin{equation}
\hat{F}_{ij}=\frac{F_{ij}-F_{\min}}{F_{\max}-F_{\min}},
\end{equation}
where $F$ represents the input feature maps, $F_{\min}$ and $F_{\max}$  denote the minimum and maximum values, respectively. In our experiments, we set $\hat{F}_{ij}>0.7$ as distinctive region. Then, we random choose $m$ pixels in the distinctive region. Around these selected pixels, $k\times k$ patches are extracted. For pixels on the edge of the feature map, the vacant neighboring pixels are filled by mirroring the image. Then, these patches are considered as kernels to be convolved with the input data.

Fig. \ref{structure} shows that the network contains multiple convolution layers. After each convolution layer, principal component analysis (PCA) is used to retain the first three principal components. The philosophy behind this behavior is that we can obtain better change detection results by fusing the multilevel features. Next, we combine the deep and shallow features which are normalized to alleviate the speckle noise. Finally, all features and the original input data are utilized to predict the changed and unchanged area via a SVM classifier. The detailed implementation steps are shown in Algorithm 1.

\begin{algorithm}[htb]
  \caption{The workflow of extracting stacked features by DPConv}
  \begin{algorithmic}[1]
    \Require
     Original data $F$, the number of patches $m$, and the depth of the network $D$.
    \Ensure
      Stacked features $F_\textrm{out}$.
     \For{each layer $d$ ($1 \leq d \leq D$)}
      \If{$(d>1)$}
       \State PCA is employed to reserve the first three principal components of $F$.
      \EndIf
      \State Find the distinctive region according to the activation value $\hat{F}_{ij}$:
      $$\hat{F}_{ij}=\frac{F_{ij}-F_{\min}}{F_{\max}-F_{\min}}$$
      \State Extract $m$ patches from the distinctive region:
      $$W=Random(m,\hat{F}_{ij}>0.7)$$
      \State Compute features by convolving $F$ with patches extracted in the previous step:
      $$DPConv=W\odot F$$
      \State Denote the features extracted in the $d^\textrm{th}$ layer as $F^{d}$
      \State If $d < D$, update the matrix $F$ by $F^d$
    \EndFor
    \State $F^1, F^2, \ldots, F^D$ are concatenated to form the final output features $F_\textrm{out}$.
 \end{algorithmic}
\end{algorithm}

\section{Experiments and Analysis}

In this section, five multitemporal SAR datasets are used to verify the effectiveness of the proposed DPDNet. An extensive study of critical parameters are presented. Next, we compare DPDNet with its variants and make a series of ablation studies. Then, the proposed DPDNet is compared with several closely related methods. Finally, we analyze the computational complexity of DPDNet to show its efficiency.

\subsection{Experimental Datasets}

To validate the effectiveness of the proposed DPDNet, we carried out experiments on five multitemporal SAR datasets. The ground truth change maps are manually annotated carefully by experts for accuracy assessment.

\begin{figure}[ht]
\centering
\includegraphics [width=3.4in]{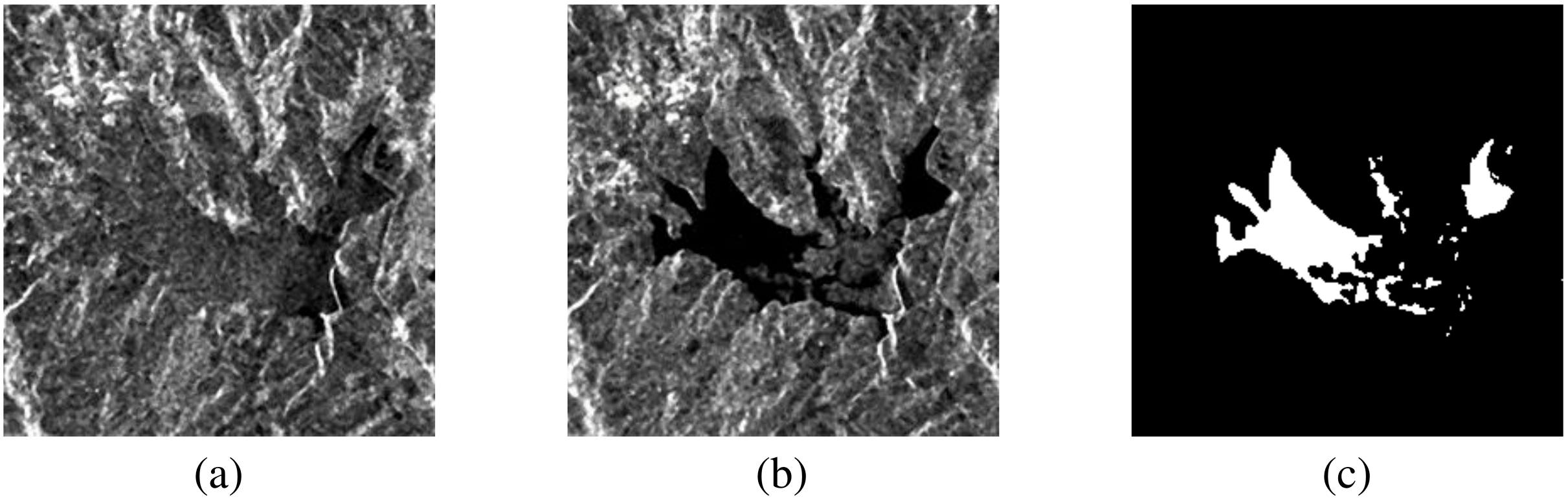}
\caption{Florence dataset. (a) Image captured on July 21, 2004. (b) Image captured on September 30, 2004. (c) Ground truth image.}
\label{Florence}
\end{figure}

The first dataset is the Florence dataset, as shown in Fig. \ref{Florence}, and it is obtained over the city of Florence, Italy. Some parts of the river have changed between the two images. The images were captured on July 21 and September 30, 2004, respectively. 

\begin{figure}[ht]
\centering
\includegraphics [width=3.4in]{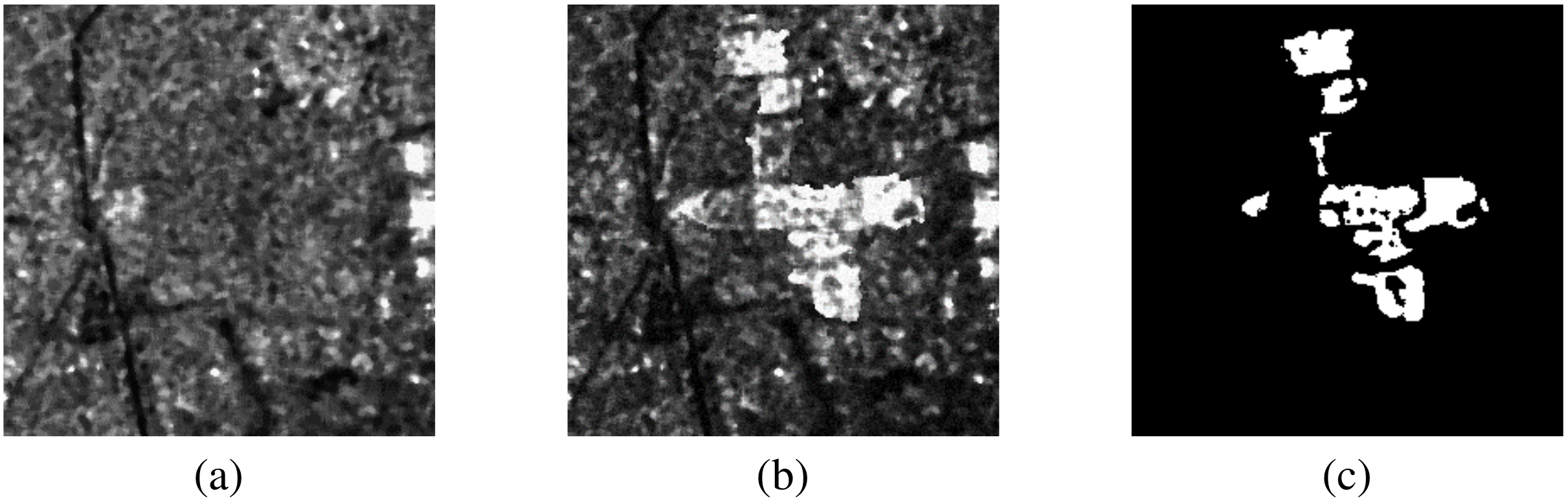}
\caption{Simulated dataset. (a) Image changed before. (b) Image changed after. (c) Ground truth image.}
\label{simulated}
\end{figure}

The second dataset is the Simulated dataset, which is a simulated dataset \cite{Gao16}.  The size of the image is 350$\times$250 pixels. The speckle noise on the Simulated dataset is rather severe. The dataset is shown in Fig. \ref{simulated}.

\begin{figure}[ht]
\centering
\includegraphics [width=3.4in]{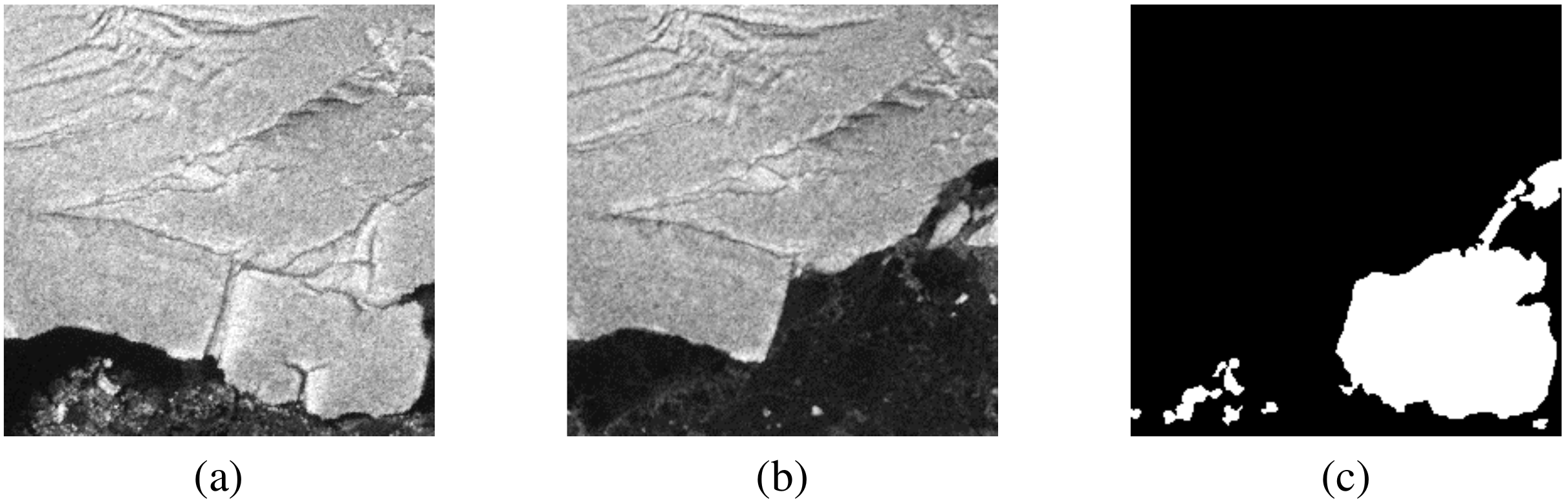}
\caption{Sulzberger dataset. (a) Image captured on March 11, 2011. (b) Image captured on March 16, 2011. (c) Ground truth image.}
\label{ice_part1}
\end{figure}

The third dataset is the Sulzberger dataset. It is cropped from a large SAR image from Sulzberger Ice Shelf. The dataset was taken by the Envisat satellite of the European Space Agency. Both images show the sea ice breakup. In March 2011, the ocean waves generated by the Tohoku Tsunami caused the Sulzberger Ice Shelf to break. To better reveal the change information, we chose one typical region. The ground truth change map is generated by experts with rich prior knowledge. The dataset is illustrated in Fig. \ref{ice_part1}.

\begin{figure}[ht]
\centering
\includegraphics [width=3.4in]{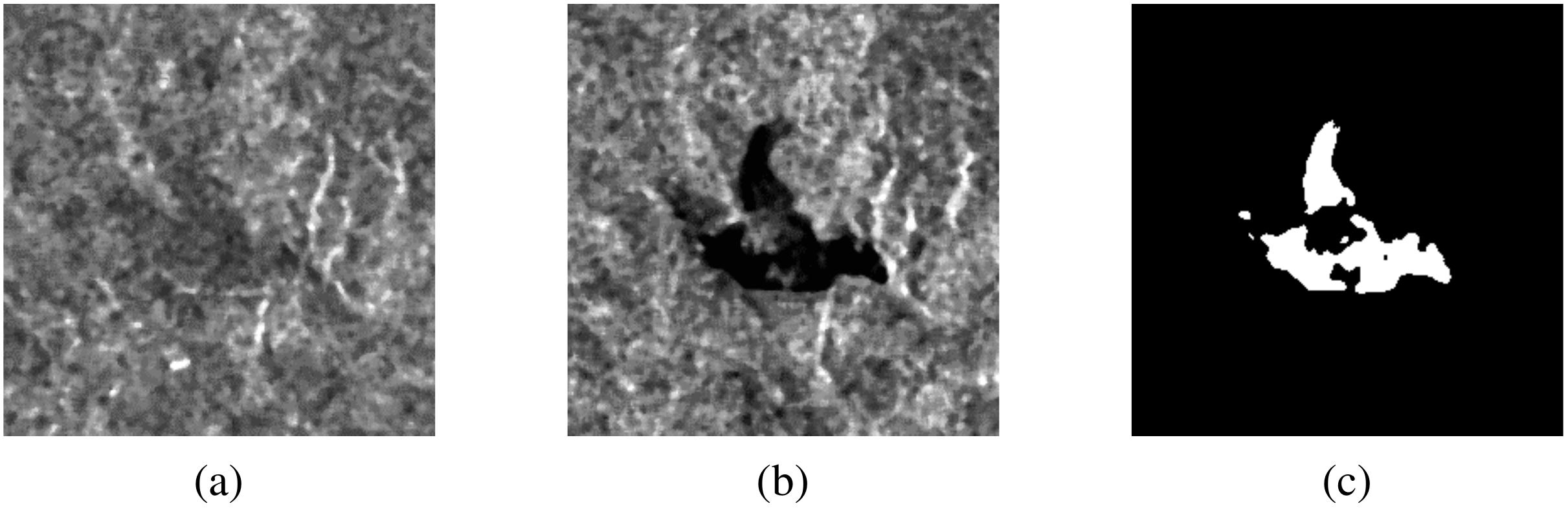}
\caption{Havana dataset. (a) Image captured in May 1997. (b) Image captured in July 1997. (c) Ground truth image.}
\label{havana}
\end{figure}

The next dataset is the Havana dataset, which is centred over the city of Havana, Cuba. Two images show the landscape changes from dry season to wet season in Havana. Two SAR images were obtained by ERS-2 satellite in May and July 1997, respectively. Some regions are flooded during the period. The Havana dataset is shown in Fig. \ref{havana}.

\begin{figure}[ht]
\centering
\includegraphics [width=3.4in]{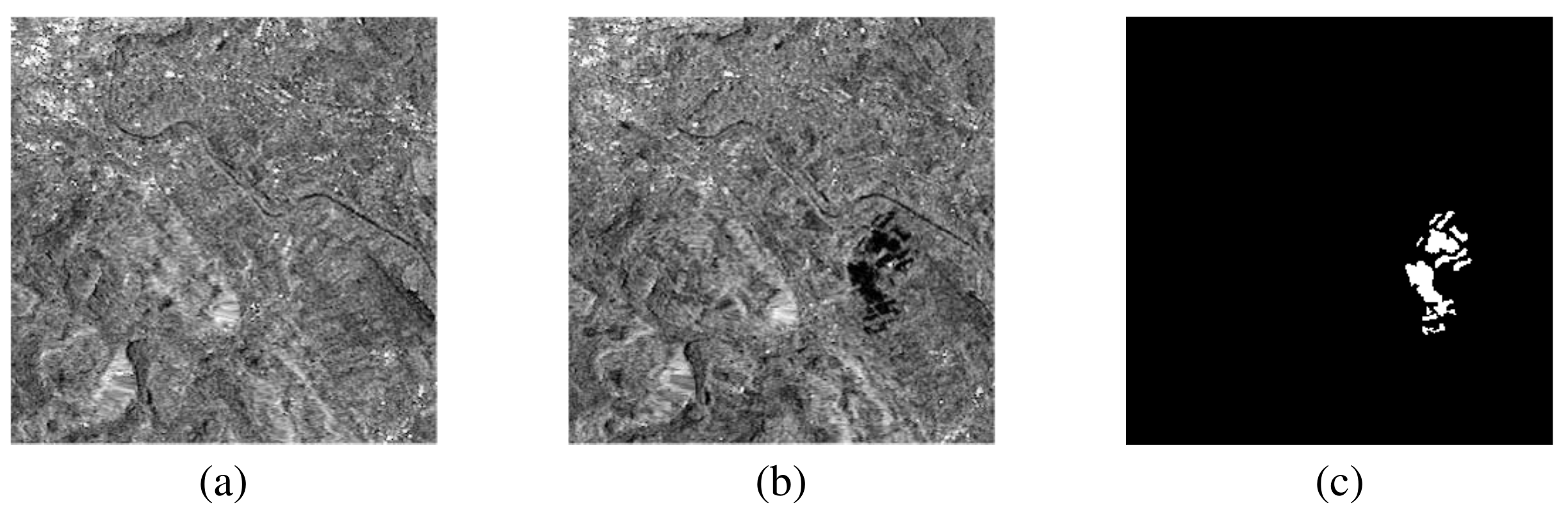}
\caption{Bern dataset. (a) Image captured in April 1997. (b) Image captured in May 1997. (c) Ground truth image.}
\label{bern}
\end{figure}

The last dataset is the Bern dataset, which is captured in the area near the city of Bern, Switzerland. Two images show the geomorphic changes after the Aare River flooded parts of the cities of Thun and Bern and the airport of Bern entirely \cite{Bazi05}. Two images were obtained by ERS-2 satellite in April and May 1997, which is illustrated in Fig. \ref{bern}.

A critical issue in change detection is the selection of evaluation criteria \cite{Gong17isprs} \cite{Gao21jstars} \cite{qu21grsl}. In this paper, false positives (FP), false negatives (FN), percentage correct classification (PCC), Kappa coefficient (KC), and F1 score (F1) are used to evaluate the change detection performance. 

1) \textbf{FP}: the number of unchanged pixels in the ground truth image but falsely classified as changed.

2) \textbf{FN}: the number of changed pixels in the ground truth image but falsely classified as unchanged.

3) \textbf{PCC}: Classification accuracy in change detection. The PCC is calculated by
\begin{equation}
\textrm{PCC} = 1-\frac{\textrm{FP}+\textrm{FN}}{N_c+N_{uc}},
\end{equation}
where $N_c$ denotes the number of changed pixels in the ground truth image, and $N_{uc}$ is the number of unchanged pixels.

4) \textbf{KC}: Kappa coefficient is used for reliability evaluation, and it is computed as:
\begin{equation}
    \textrm{KC} = \frac{\textrm{PCC}-\textrm{PRE}}
    {1 - \textrm{PRE}},
\end{equation}
\begin{equation}
    \textrm{PRE}=\frac{
    (N_c+\textrm{FP}-\textrm{FN})\times N_c +
    (N_{uc}+\textrm{FN}-\textrm{FP})\times N_{uc}}{
    (N_c + N_{uc}) \times (N_c + N_{uc})},
\end{equation}

5) \textbf{F1}: F1 score is the harmonic mean of precision and sensitivity, and it is computed as:
\begin{equation}
\textrm{F1} = \frac{\textrm{2TP}}{\textrm{2TP}+\textrm{FP}+\textrm{FN}},
\end{equation}
where \textrm{TP} is the number of changed pixels in the ground truth image and truly classified as changed.

\subsection{Analysis of the Important Parameters of DPDNet}

\begin{figure}[ht]
\centering
\includegraphics [width=3.2in]{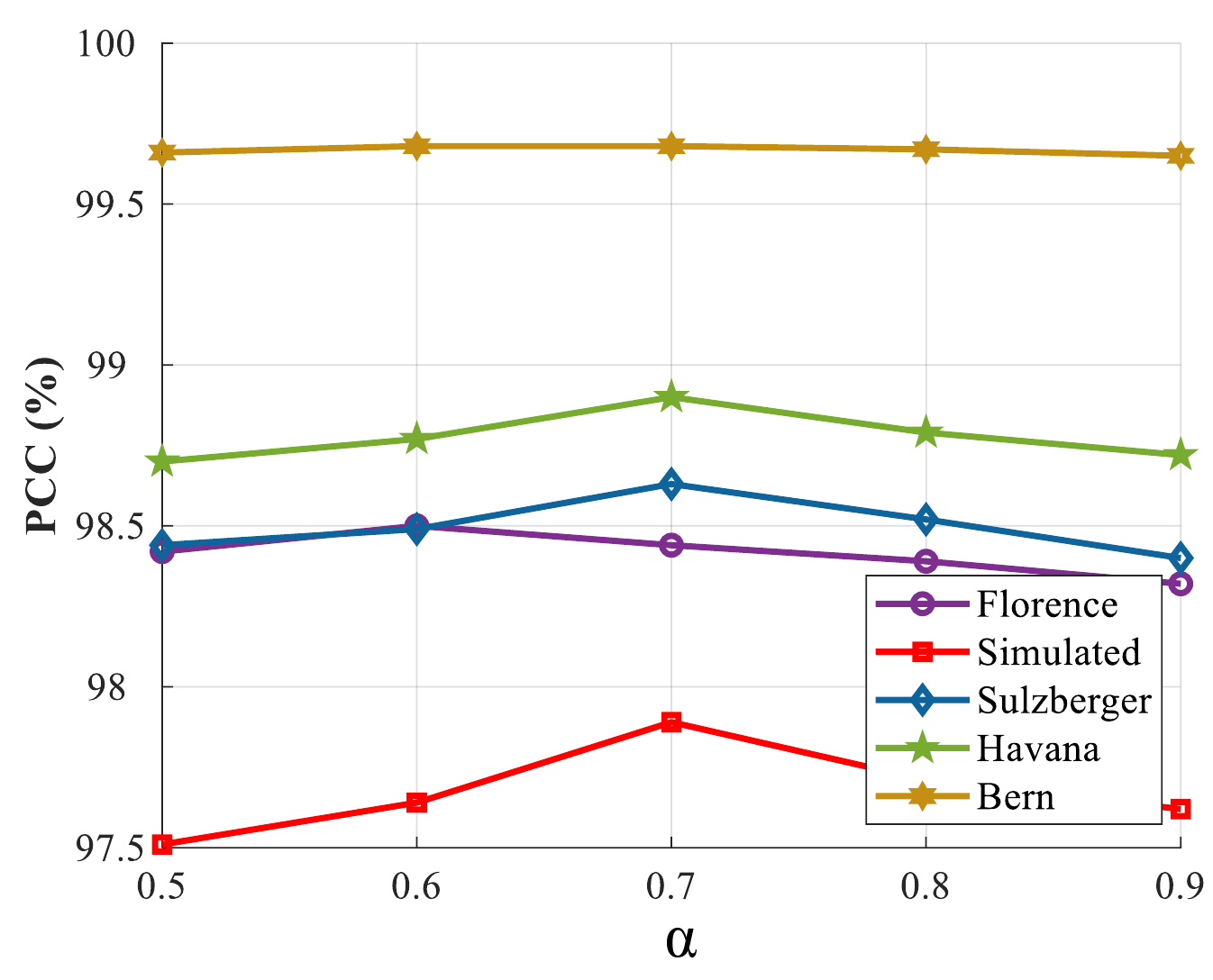}
\caption{Relationship between the classification accuracy and parameter $\alpha$.}
\label{alpha}
\end{figure}

\textbf{Analysis of the parameter $\alpha$.} In the proposed DPDNet, an important branch is used to clean the label noise. The critical parameter $\alpha$ is used to balance the contribution between the initial label itself and the labels received from its neighbors. We evaluate the change detection performance by taking $\alpha$ = 0.5, 0.6, 0.7, 0.8, and 0.9. Fig. \ref{alpha} shows the quantitative analysis result of the parameter $\alpha$. It can be observed that when $\alpha = 0.6$, the PCC obtains the best value on the Florence datasets, while on the Simulated, Sulzberger, Havana and Bern datasets, the result achieves the best when $\alpha = 0.7$. Since different datasets have different spatial structures in the label propagation, the contributions between the initial label and the label received from its neighbors are different. Hence, we set $\alpha$ to 0.6 on the Florence datasets, and set $\alpha$ to 0.7 on the Simulated, Sulzberger, Havana and Bern datasets in our following experiments.

\begin{figure}[ht]
\centering
\includegraphics [width=3.2in]{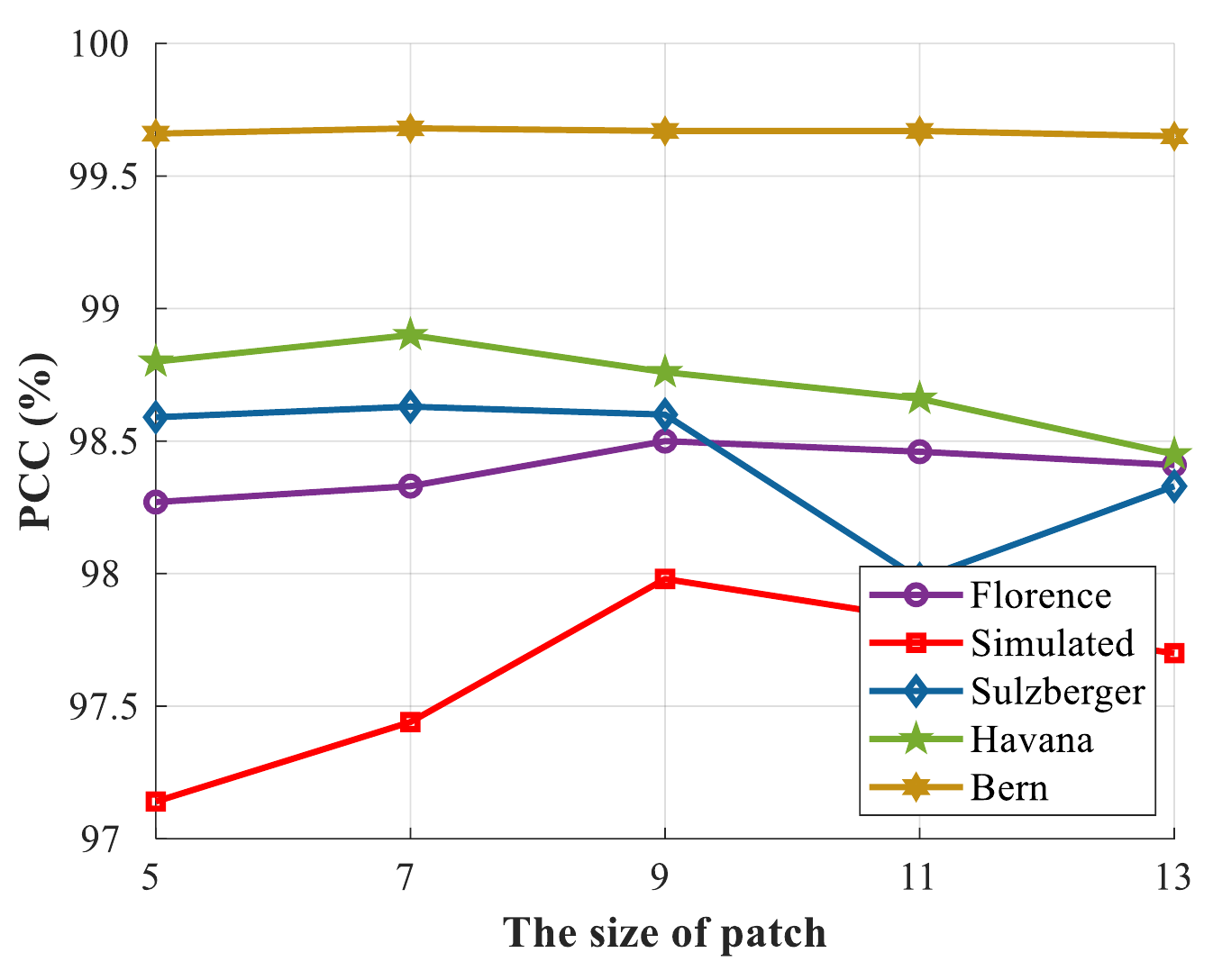}
\caption{Relationship between PCC and the size of patch.}
\label{patchsize}
\end{figure}

\textbf{Analysis of the image patch size.} As we all know, the contextual information plays an important role in alleviating the speckle noise. However, the patch size controls the contextual information contained in the training samples, and therefore we design experiments to test the parameter $w$. Here, $w$ denotes the size of the patches used as the training sample. We evaluate the change detection performance by taking $w$ = 5, 7, 9, 11, and 13. As shown in Fig. \ref{patchsize}. We can see that the PCC value is not satisfying when $w=5$, because such small patches can hardly extract the contextual information effectively. On the Sulzberger, Havana, and Bern datasets, the DPDNet achieves the best result when $w=7$. On the Florence and Simulated datasets, $w=9$ obtains the best PCC values. It indicates that different datasets contain different contextual information and require different size of receptive fields for feature extraction. Therefore, in the following implementation, we set $w=7$ on the Sulzberger, Havana and Bern datasets and set $w=9$ on the Florence and Simulated datasets.

\begin{figure}[ht]
\centering
\includegraphics [width=3.2in]{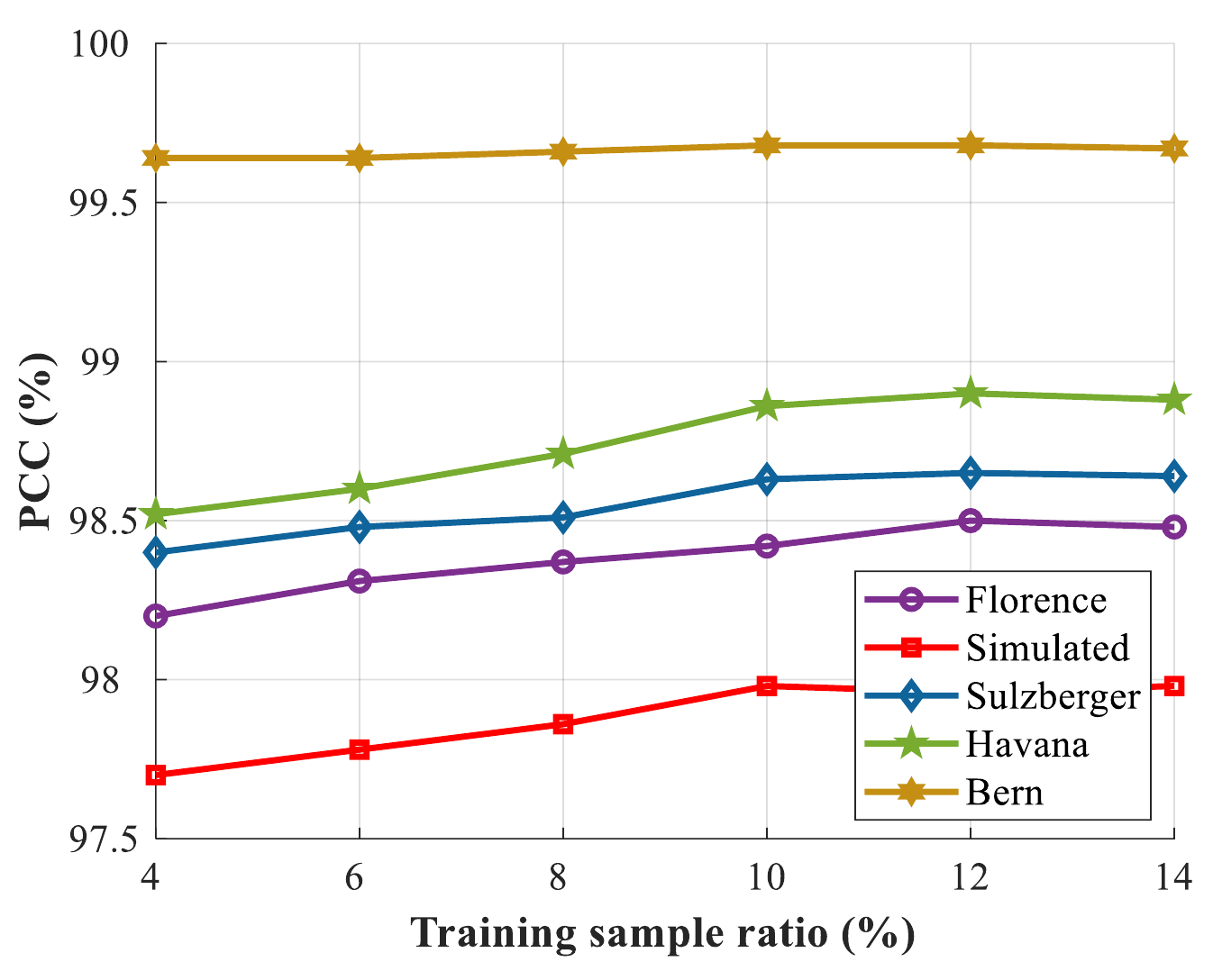}
\caption{Relationship between the  change detection accuracy and the number of training samples.}
\label{trainratio}
\end{figure}

\textbf{Analysis of the training sample numbers.} It is evident that the number of training samples has a great impact on the change detection performance. Hence, we investigate the relationship between PCC value and the training sample numbers on different datasets. As mentioned before, the training samples are randomly selected from the preclassification results by the clustering algorithm. We randomly selected 4\%, 6\%, 8\%, 10\%, 12\% and 14\% pixels as training samples. The change detection performance on different datasets by using different training sample numbers are shown in Fig. \ref{trainratio}. From the curves in the figure, it can be seen that there is a sharp increase in PCC value when the training sample number ranges from 4\% to 10\%. On Florence, Sulzberger, Havana, and Bern datasets, the best result is achieved when the training sample ratio is 12\%. After that, the PCC value tends to be stable even the number of training samples increases. Hence, we select 12\% pixels from the preclassification result as training samples.

\begin{figure}[ht]
\centering
\includegraphics [width=3.2in]{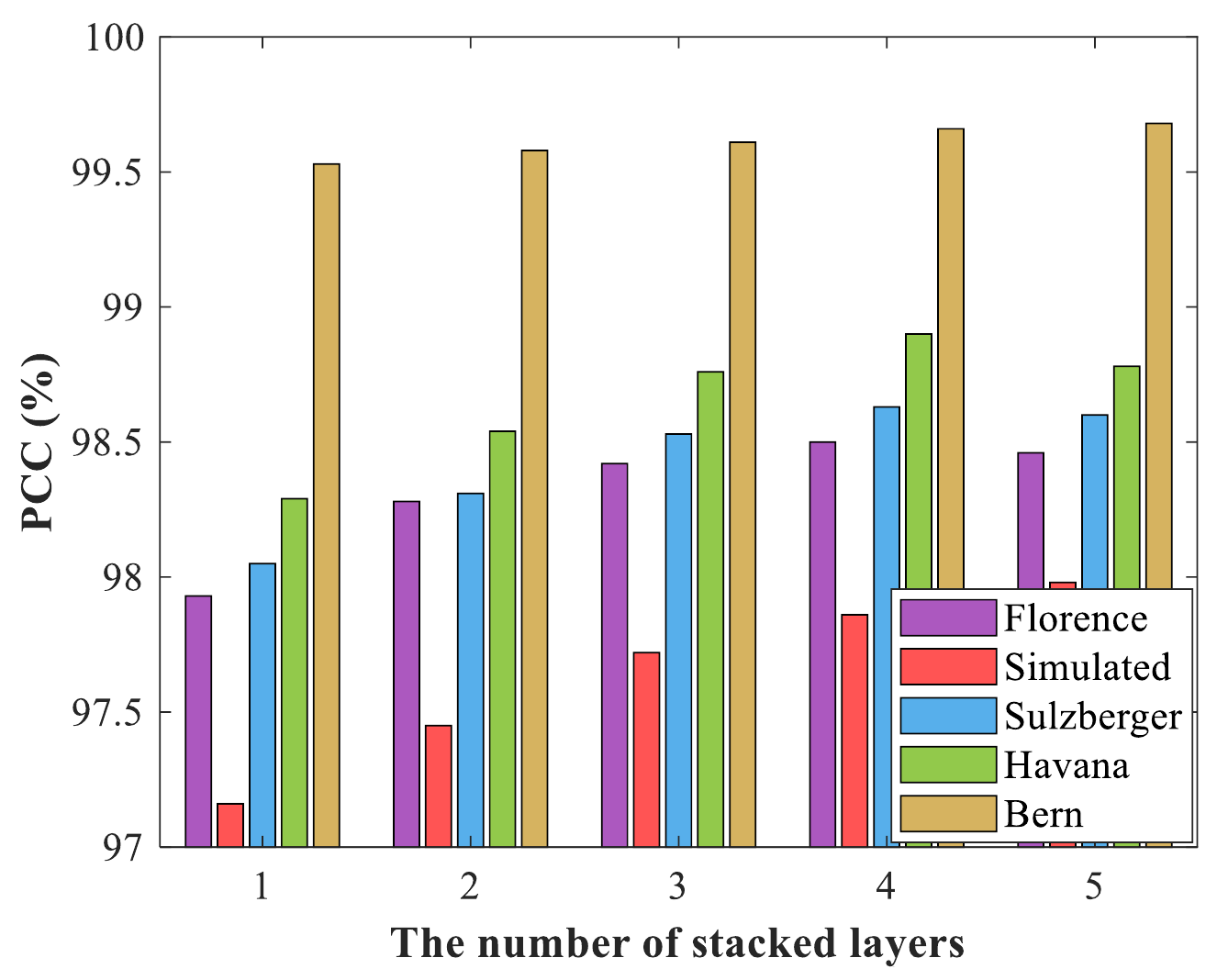}
\caption{Relationship between PCC and the stacked layer number.}
\label{layernumber}
\end{figure}

\begin{figure}[ht]
\centering
\includegraphics [width=3.3in]{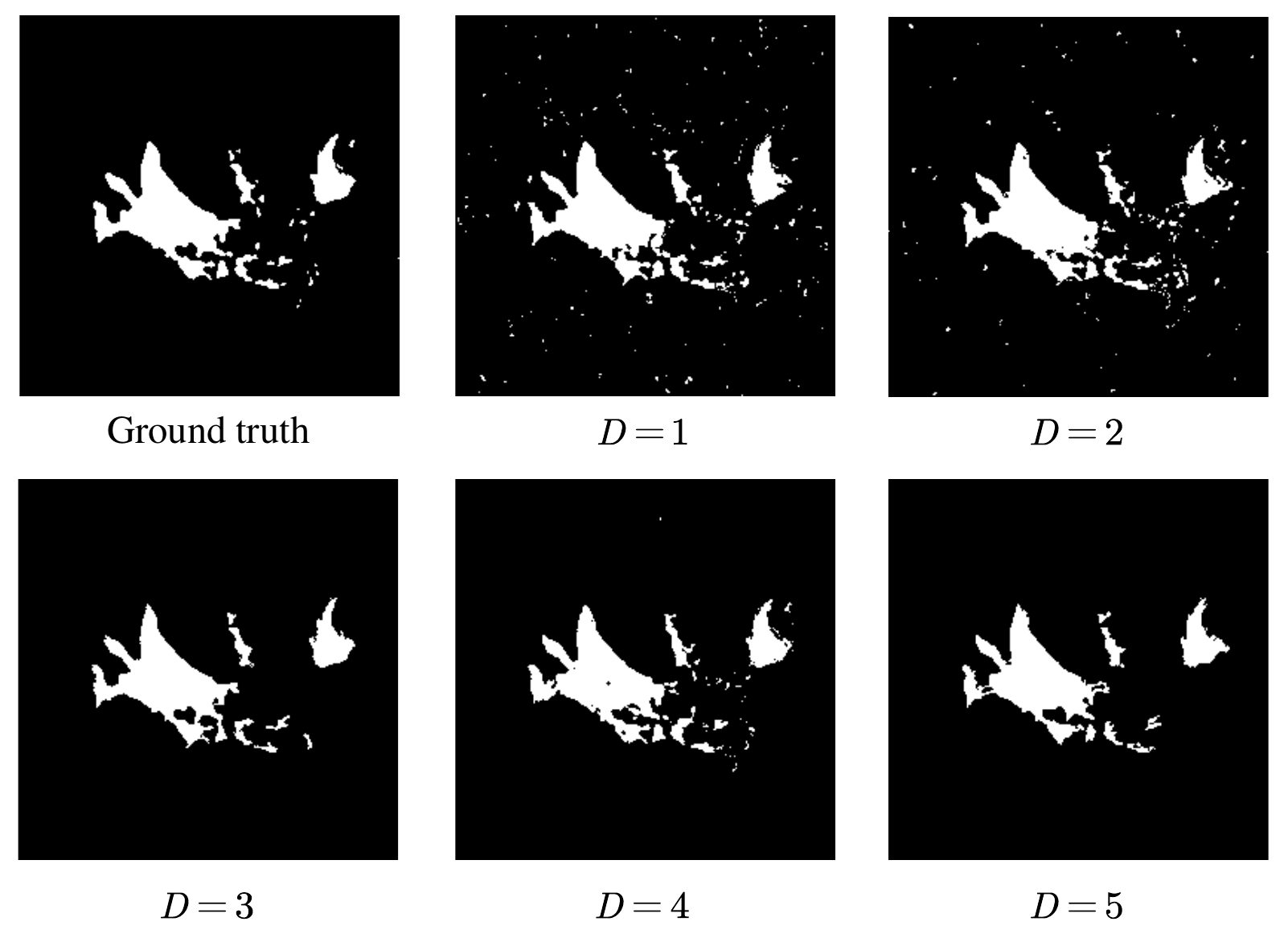}
\caption{Change detection results on Florence dataset by different stacked features.}
\label{layernumber_Florence}
\end{figure}

\begin{figure}[ht]
\centering
\includegraphics [width=3.3in]{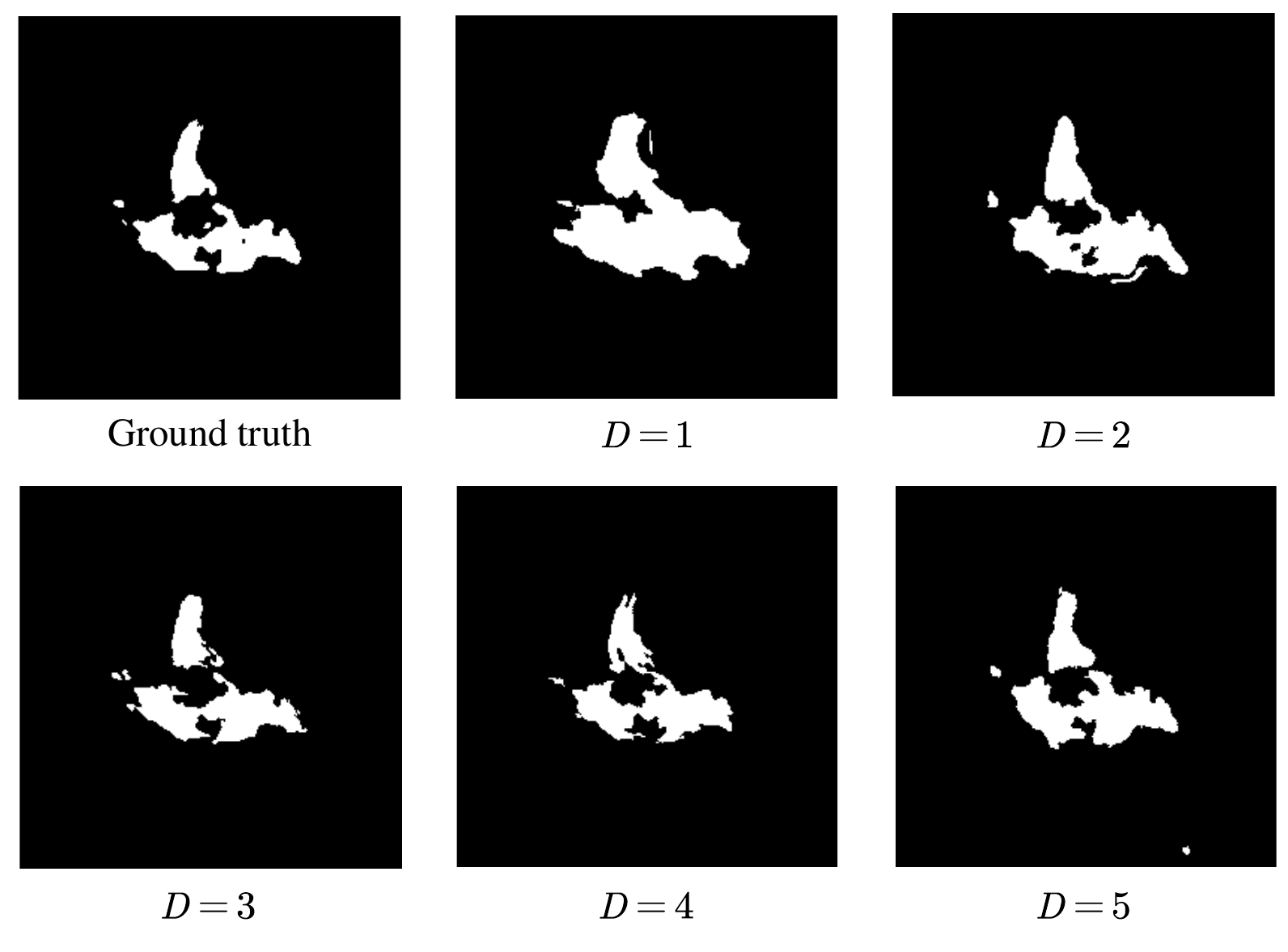}
\caption{Change detection results on Havana dataset by different the number of stacked features.}
\label{layernumber_Seoul}
\end{figure}

\textbf{Analysis of the stacked layer numbers.} The main objective of this study is to use distinctive patches extracted from the original images as convolution kernels for feature extraction. In this paper, we use multiple convolutional layers to extract shallow and deep features. Therefore, the number of convolutional layers affects the number of stacked features, which has a great impact on the change detection results. In this experiment, we explore the relationship between the number of stacked layers $D$ and the change detection results. The experimental results are illustrated in Fig. \ref{layernumber}. As can be observed, when $D=4$, the PCC values reach the best on the Florence, Sulzerberg, and Havana datasets. On the Simulated and Bern dataset, the best performance is achieved when $D=5$. It is owing to the reason that different datasets have different texture details. With the increase of stacked layers, better feature representations can be captured on different regions. The visualized results with different number of stacked layers on the Florence and Havana datasets are shown in Fig. \ref{layernumber_Florence} and Fig. \ref{layernumber_Seoul}, respectively. As can be seen, with the increase of stacked layers, the change map contains less noise regions. Therefore, $D=5$ is used on the Simulated and Bern dataset, $D=4$ is employed on other datasets. In real applications, the value of $D$ should be adjusted according to the texture or geometry characteristics of each dataset.

\subsection{Verification of Distinctive Patch Convolution effectiveness}
\label{sec1}

\begin{figure}[ht]
\centering
\includegraphics [width=3.4in]{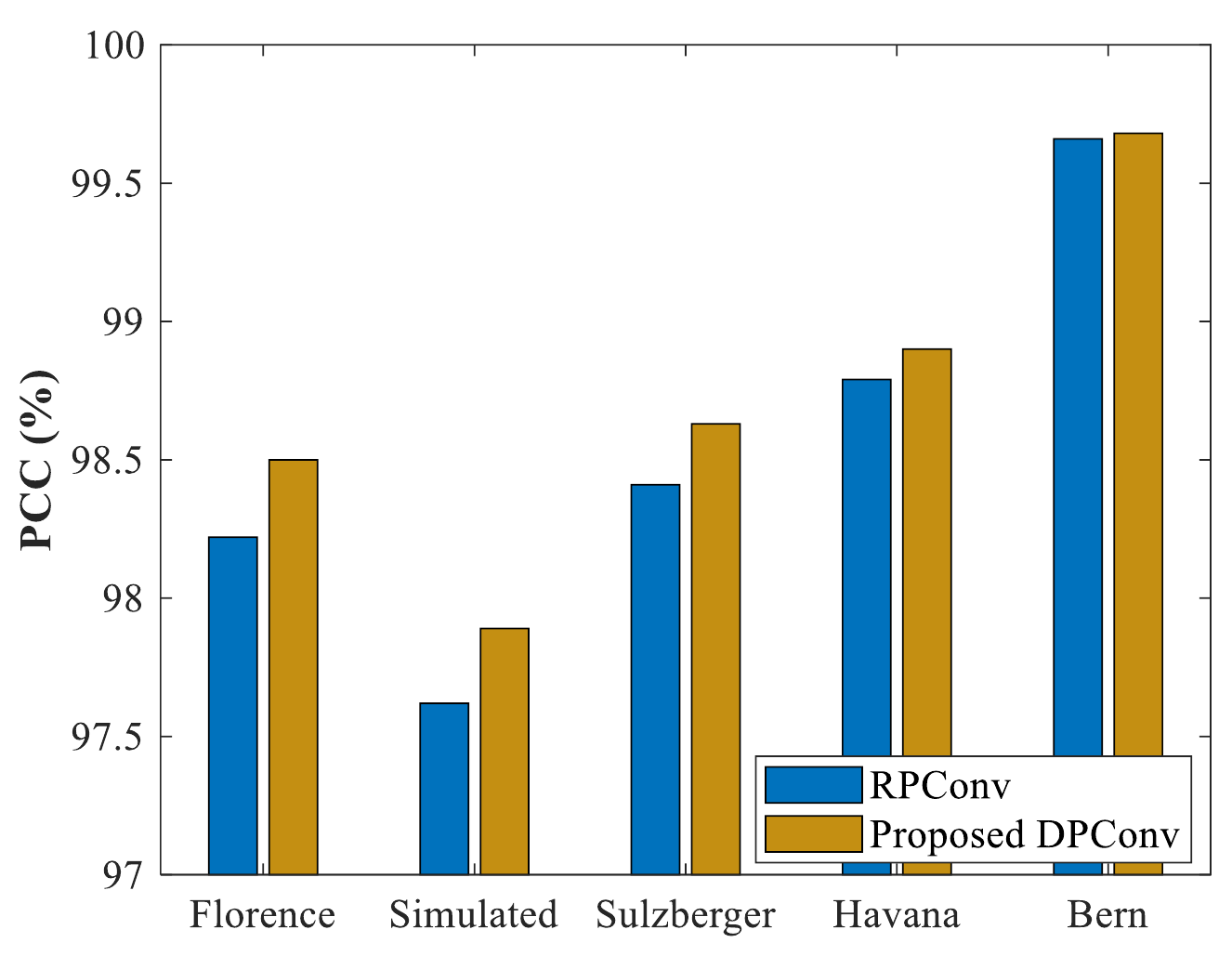}
\caption{The comparison between distinctive patch convolution with random patch convolution.}
\label{compare_with_rpnet}
\end{figure}

\textbf{Distinctive Patch Convolution versus Random Patch Convolution.}To alleviate the speckle noise, our method introduces the distinctive patch convolution (DPConv) to extract the desired features. In most existing works, a great number of training samples are needed to optimize the parameters of the network, which is time-consuming. To make up this shortcoming, Xu {\it et al.} \cite{Xu18} proposed  random patch convolution (RPConv) in which random patches taken from the image are used as convolution kernels. However, the randomness of RPConv risks the model's stability. Compared with RPNet, the proposed DPConv pinpoints the most non-trivial units of the image and offers better feature representation upon sampling position. We compare the models using DPConv with RPConv for feature extraction. As illustrated in Fig. \ref{compare_with_rpnet}, in most cases, the standard DPDNet offers a better performance. Meanwhile, the performance of RPConv is not stable, and it achieves relatively good change detection results when it takes samples from representative locations.

\begin{figure}[ht]
\centering
\includegraphics [width=3.4in]{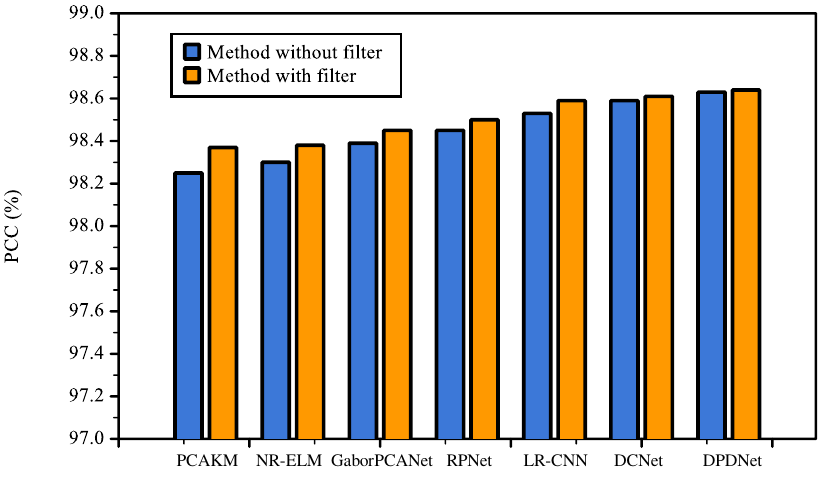}
\caption{Effect of filtering operations on Sulzberger dataset.}
\label{filter}
\end{figure}

\textbf{Distinctive Patch Convolution versus Non-local filter.} To verify the performance of the proposed DPConv for speckle noise suppression, the non-local means filter \cite{Su15isprs} was used for data preprocessing for all methods. As shown in Fig. \ref{filter}, the filtering operation is performed on the Sulzberger dataset for these methods, and PCC value is utilized as the evaluation metric. It is obvious that non-local filter reduces the speckle noise to some extent and improves the change detection performance for most of the methods. As for DPDNet, the performance is also improved after using non-local filter for data preprocessing, but the improvement was smaller compared to the other methods. This is because it provides an end-to-end way to reduce the speckle noise via the strong learning capability of the network. Furthermore, the evaluation metric shows that DPDNet still outperforms the other methods without the addition of filter, which also proves that it is beneficial in speckle noise suppression.

\begin{table*}[htbp]
\centering
\begin{center}
\renewcommand\arraystretch{1.60}
\renewcommand\tabcolsep{5.0pt}
\caption{Ablation Studies on Four Datasets}
\begin{tabular}{c|ccc|ccccc}
\toprule
\multirow{2}{*}{\#} &
\multirow{2}{*}{~DPConv~}&
\multirow{2}{*}{~RPConv~}&
\multirow{2}{*}{~RLPA~} &
\multicolumn{5}{c}{PCC (\%) on different datasets} \\ \cline{5-9}
\multicolumn{1}{c|}{}& \multicolumn{3}{c|}{} & ~~Florence~~ & ~~Simulated~~ & ~~Sulzberger~~ & ~~Havana~~ & ~~Bern~~ \\ \midrule
~~~ 1 ~~~  & \XSolidBrush & \XSolidBrush & \XSolidBrush & 98.02 & 97.43 & 98.22 & 98.37 &99.60 \\
2  & \Checkmark   & \XSolidBrush & \XSolidBrush & 98.39 & 97.71 & 98.48 & 98.72 & 99.65\\
3  & \XSolidBrush & \Checkmark   & \XSolidBrush & 98.14 & 97.57 & 98.34 & 98.56 & 99.63\\
4  & \XSolidBrush & \XSolidBrush & \Checkmark   & 98.13 & 97.54 & 98.39 & 98.50 & 98.62\\
5  & \XSolidBrush & \Checkmark   & \Checkmark   & 98.21 & 97.66 & 98.47 & 98.71 &99.65\\
6  & \Checkmark   & \XSolidBrush & \Checkmark   & \textbf{98.50} & \textbf{97.98} & \textbf{98.63} & \textbf{98.90} &\textbf{99.68}\\
\bottomrule
\end{tabular}
\label{table_ablation}
\end{center}
\end{table*}

\subsection{Ablation Study}

We further verify the effectiveness of both branches of the proposed DPDNet on five datasets. Note that to make a fair comparison, we have designed a Baseline (\#1), which is completely consistent with the structure of the branch using DPConv in DPDNet, including the size of the convolution kernel and the number of layers, except that DPConv is replaced by the traditional convolution.

The Baseline (\#1) has the same number of training epochs and training samples as DPDNet. However, as reported in Table \ref{table_ablation}, the performance of $Baseline$ is unsatisfying since the number of training samples is quite limited. When we use the distinctive patch extracted from the image as convolution kernels, it reduces the dependence of the network parameters and achieves better results (\#1 vs \#3). Meanwhile, the RPConv will risk the stability of the network. For instance, when the kernels are selected from the smooth-textured background, the parameters of the convolution kernels are similar, and the extracted features cannot be representative. Hence, the proposed DPConv can effectively select kernels from the distinctive region, so as to reduce the influence of randomness. The results of \#2 and \#3 in Table \ref{table_ablation} demonstrate the effectiveness of the proposed DPConv in feature representation.

Furthermore, label noise is introduced into change detection. To handle the problem, we use RLPA to clean the label noise hidden in the preclassification results. To evaluate the effectiveness of label noise cleaning, we make extensive experiments  (\#4 vs \#1, \#6 vs \#2, \#5 vs \#3). It is evident that when we introduce a label noise cleaning branch, the change detection performance always significantly improved.  Moreover, it can be observed that the proposed DPDNet (\#6) not only enhances the feature representation by using DPConv, but also improves the classification performance by alleviating the label noise in preclassification.

\subsection{Change Detection Results on Five SAR Datasets}

To validate the performance of the proposed DPDNet, six state-of-the-art methods, including PCAKM \cite{Celik09}, NR-ELM \cite{Gao16_jars}, GaborPCANet \cite{Gao16_grsl}, RPNet \cite{Xu18}, LR-CNN \cite{Liu19} and DCNet \cite{gao19jstars} are used as the comparison methods.
For PCAKM, the contextual information is analyzed by PCA, and the features are handled $k$-means clustering. NR-ELM utilizes extreme learning machine (ELM) \cite{huang04ijcnn} as the classifier. GaborPCANet is composed of cascaded PCA layer, binary hashing layer, and block-wise histogram generation. RPNet uses random patches taken from the original image as convolution kernels to extract shallow and deep concolutional features. LR-CNN is a deep learning based method with local spatial restrictions. In DCNet, a channel weighting-based module is designed for feature enhancement. Among these methods, RPNet, LR-CNN, and DCNet are deep learning-based methods. It should be noted that the RPNet is originally designed for hyperspectral image classification, and we made some adjustments so that it can be applied to change detection tasks.

\begin{figure}[]
\centering
\includegraphics [width=3.3in]{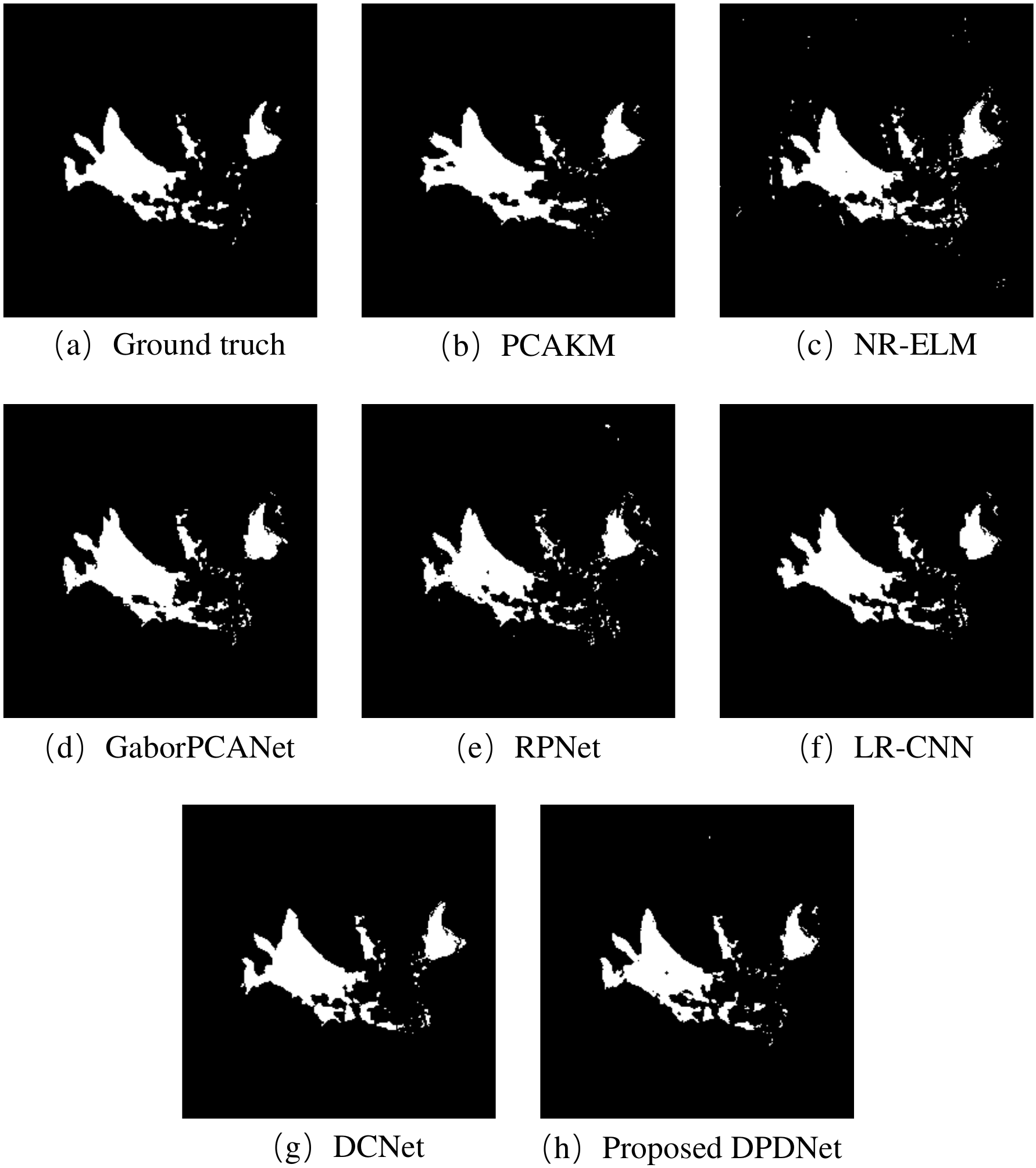}
\caption{Comparisons of the change maps generated by different methods on the Florence dataset. (a) Ground truth image. (b) Result by PCAKM \cite{Celik09}. (c) Result by NR-ELM \cite{Gao16_jars}. (d) Result by GaborPCANet \cite{Gao16_grsl}. (E) Result by RPNet \cite{Xu18}. (F) Result by LR-CNN \cite{Liu19}. (g) Result by DCNet \cite{gao19jstars}. (h) Result by the proposed DPDNet.}
\label{result_Florence}
\end{figure}

\begin{table}[]
\centering
\caption{Change detection results of different methods on the Florence dataset}
\renewcommand\arraystretch{1.5}
\setlength{\tabcolsep}{1.0mm}{
\begin{tabular}{c|ccccc}
\toprule
Methods & ~ FP ~ & ~ FN ~ & ~ PCC(\%) ~ & KC(\%) ~ & F1(\%) \\ \midrule
PCAKM \cite{Celik09} & 1289 & 672 & 97.01 & 78.59 & 82.20 \\
NR-ELM \cite{Gao16_jars} & 1203 & 590  & 97.26 & 86.03 & 83.34 \\
GaborPCANet \cite{Gao16_grsl} & 1055 & 564  & 97.53 & 86.19 & 84.86\\
RPNet \cite{Xu18} & 987 & 536  & 97.68 & 86.24 & 85.07\\
LR-CNN \cite{Liu19} & 882 & 491  & 97.90 & \textbf{86.32} & 86.78\\
DCNet \cite{gao19jstars} & 805 & 410  & 98.15 & 86.31 & 88.06\\
Proposed DPDNet & \textbf{686} & \textbf{298}  & \textbf{98.50} & \textbf{86.32} &\textbf{89.88} \\
\bottomrule
\end{tabular}
}
\label{table_Florence}
\end{table}

\textbf{Results on the Florence dataset.} Fig. \ref{result_Florence} shows the experimental results on the Florence dataset, while Table \ref{table_Florence} lists the corresponding quantitative metrics. From Table \ref{table_Florence}, we can see that, compared with DPDNet, other methods suffer from high FP values resulting in many noisy spots in the change map as shown in Fig. \ref{result_Florence}. Since other learning-based methods do not take the label noise into account, and the training samples with errors mislead the learning process. In addition, the FN value of DPDNet is the lowest, which means that the proposed model does not ignore some subtle change pixels. Among these methods, the proposed DPDNet achieved the best PCC, KC, and F1 values. The results on the Florence dataset demonstrate that label cleaning is necessary.  The comparisons demonstrate the superior performance of the proposed DPDNet on the Florence dataset.

\begin{figure}[]
\centering
\includegraphics [width=3.3in]{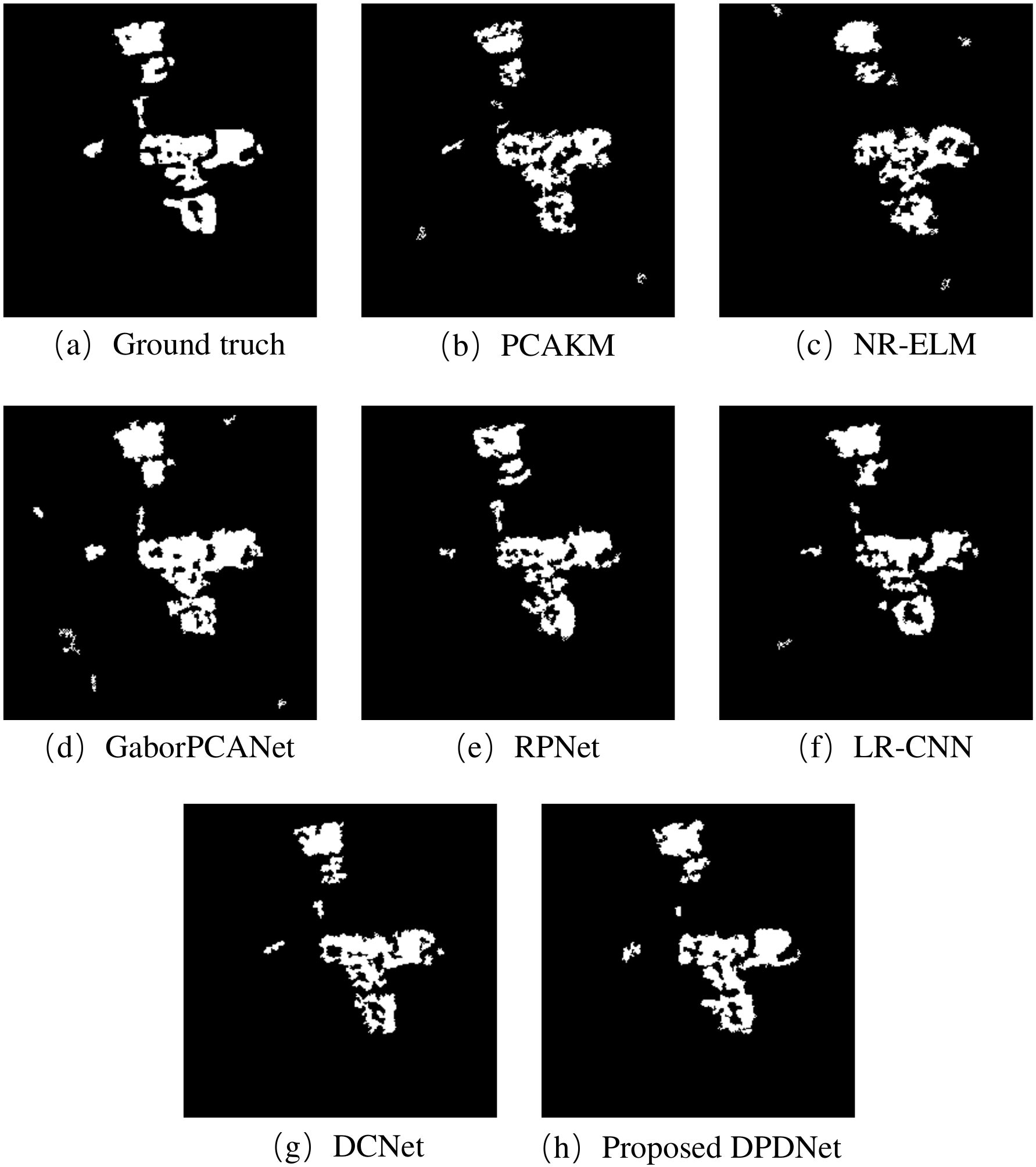}
\caption{Comparisons of the change maps generated by different methods on the Simulated dataset. (a) Ground truth image. (b) Result by PCAKM \cite{Celik09}. (c) Result by NR-ELM \cite{Gao16_jars}. (d) Result by GaborPCANet \cite{Gao16_grsl}. (E) Result by RPNet \cite{Xu18}. (F) Result by LR-CNN \cite{Liu19}. (g) Result by DCNet \cite{gao19jstars}. (h) Result by the proposed DPDNet.}
\label{result_Simulated}
\end{figure}

\begin{table}[]
\centering
\caption{Change detection results of different methods on the Simulated dataset}
\renewcommand\arraystretch{1.5}
\setlength{\tabcolsep}{1.0mm}{
\begin{tabular}{c|ccccc}
\toprule
Methods & ~ FP ~ & ~ FN ~ & ~ PCC(\%) ~ & KC(\%)  ~ & F1(\%) \\ \midrule
PCAKM \cite{Celik09} & 680 & 1247  & 97.06 & 74.63 & 76.19\\
NR-ELM \cite{Gao16_jars} & 903 & 1286  & 96.66 & 71.77 & 73.55\\
GaborPCANet \cite{Gao16_grsl} & 1206 & \textbf{507}  & 97.39 & 80.30 & 81.70\\
RPNet \cite{Xu18} & 641 & 831  & 97.75 & 81.42 & 82.62\\
LR-CNN \cite{Liu19} & 523 & 906  & 97.82 & 81.57 & 82.74\\
DCNet \cite{gao19jstars} & \textbf{420} & 979  & 97.87 & 81.60 & 82.73\\
Proposed DPDNet & 551 & 771  & \textbf{97.98} & \textbf{83.26} & \textbf{84.34}\\
\bottomrule
\end{tabular}
}
\label{table_Simulated}
\end{table}

\textbf{Results on the Simulated dataset.} Fig. \ref{result_Simulated} presents the change maps on the Simulated dataset. The corresponding quantitative metrics are shown in Table \ref{table_Simulated}. In addition to the complex spatial structure, the Simulated dataset is severely interfered by speckle noise. Thus, it is rather challenging to identify changed pixels accurately on this dataset. The results of NR-ELM and GaborPCANet are polluted with noisy regions. Therefore, NR-ELM and GaborPCANet suffer from high FP values. Generally speaking, the deep learning-based models improve the change detection performance with a clear margin. In addition, the proposed DPDNet performs favorably against other methods on this dataset. The comparison shows that the DPDNet is powerful in noise suppression on the Simulated dataset.

\begin{figure}[]
\centering
\includegraphics [width=3.3in]{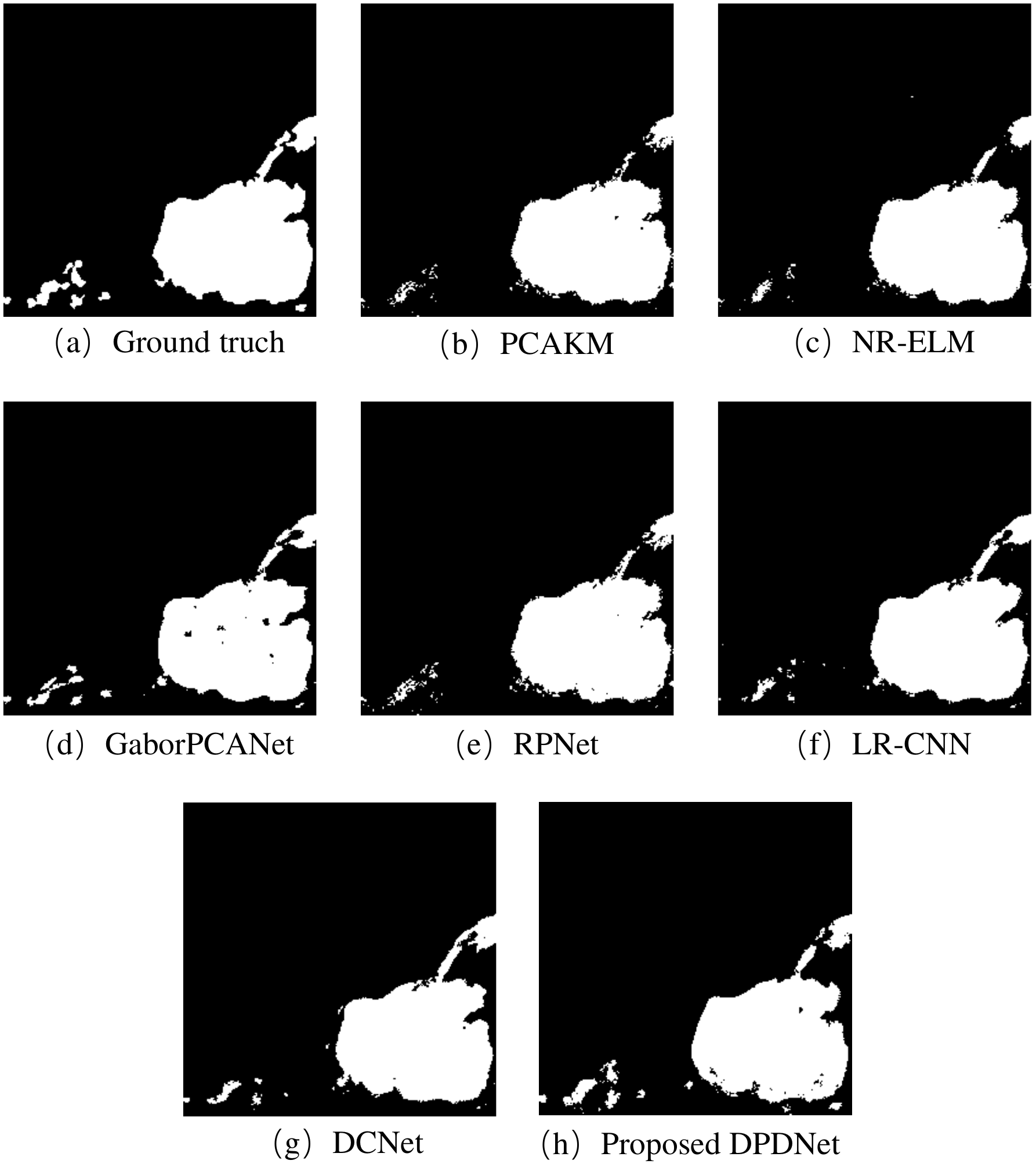}
\caption{Comparisons of the change maps generated by different methods on the Sulzberger dataset. (a) Ground truth image. (b) Result by PCAKM \cite{Celik09}. (c) Result by NR-ELM \cite{Gao16_jars}. (d) Result by GaborPCANet \cite{Gao16_grsl}. (E) Result by RPNet \cite{Xu18}. (F) Result by LR-CNN \cite{Liu19}. (g) Result by DCNet \cite{gao19jstars}. (h) Result by the proposed DPDNet.}
\label{result_ice}
\end{figure}

\begin{table}[]
\centering
\caption{Change detection results of different methods on the Sulzberger dataset}
\renewcommand\arraystretch{1.5}
\setlength{\tabcolsep}{1.0mm}{
\begin{tabular}{c|ccccc}
\toprule
Methods & ~ FP ~ & ~ FN ~ & ~ PCC(\%) ~ & KC(\%) ~ & F1(\%)\\ \midrule
PCAKM \cite{Celik09} & 144 & 998 & 98.25 & 93.91 & 94.97\\
NR-ELM \cite{Gao16_jars} & \textbf{74} & 1037 & 98.30 & 94.07 & 95.09\\
GaborPCANet \cite{Gao16_grsl} & 203 & 850 & 98.39 & 94.44 & 95.41\\
RPNet \cite{Xu18} & 184 & 831  & 98.45 & 94.64 &95.58\\
LR-CNN \cite{Liu19} & 151 & 810  & 98.53 & 94.92 &95.81\\
DCNet \cite{gao19jstars} & 167 & 756 & 98.59 & 95.13 &95.99\\
Proposed DPDNet & 240 & \textbf{659} & \textbf{98.63} & \textbf{95.29} & \textbf{96.12}\\
\bottomrule
\end{tabular}
}
\label{table_Sulzberger}
\end{table}

\textbf{Results on the Sulzberger dataset.} Fig. \ref{result_ice} shows the change detection results on the Sulzberger dataset. From visual inspection, it can be seen that most methods missed some subtle changed areas, which generate higher FN values in Table \ref{table_Sulzberger}. Moreover, deep learning-based methods (LR-CNN, RPNet, and DCNet) generally obtain better performance. Furthermore, because of the stacked DPConv features and label cleaning, the proposed DPDNet obtains the best PCC, KC, and F1 values. Overall, the proposed DPDNet outperforms in terms of PCC, KC, and F1 values on the Sulzberger dataset.

\begin{figure}[]
\centering
\includegraphics [width=3.3in]{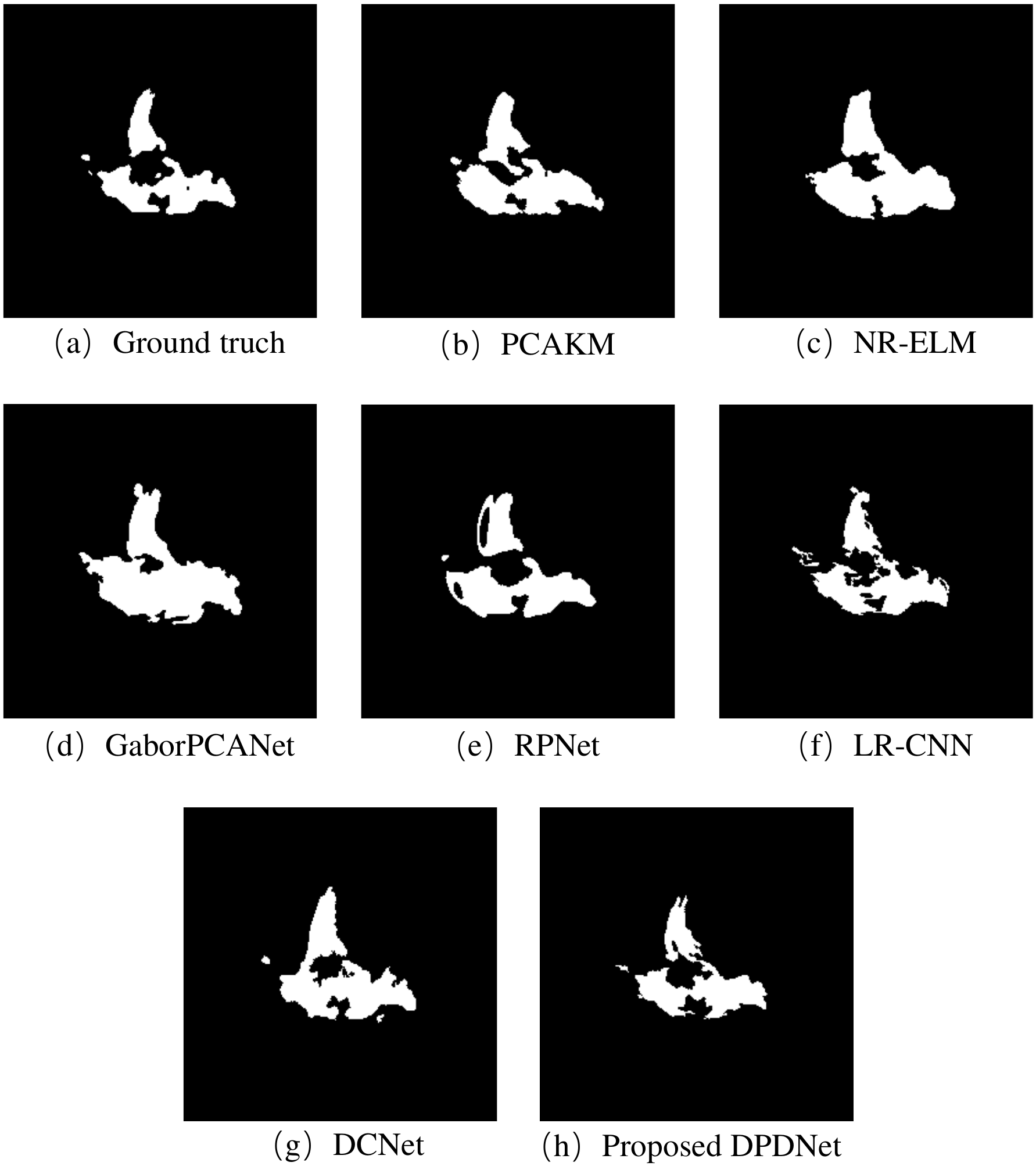}
\caption{Comparisons of the change maps generated by different methods on the Havana dataset. (a) Ground truth image. (b) Result by PCAKM \cite{Celik09}. (c) Result by NR-ELM \cite{Gao16_jars}. (d) Result by GaborPCANet \cite{Gao16_grsl}. (E) Result by RPNet \cite{Xu18}. (F) Result by LR-CNN \cite{Liu19}. (g) Result by DCNet \cite{gao19jstars}. (h) Result by the proposed DPDNet.}
\label{result_Havana}
\end{figure}

\begin{table}[]
\centering
\caption{Change detection results of different methods on the Havana dataset}
\renewcommand\arraystretch{1.5}
\setlength{\tabcolsep}{1.0mm}{
\begin{tabular}{c|ccccc}
\toprule
Methods & ~ FP ~ & ~ FN ~ & ~ PCC(\%) ~ & KC(\%) ~ & F1(\%)\\ \midrule
PCAKM \cite{Celik09} & 1077 & 119 & 98.18 & 83.98 & 84.93\\
NR-ELM \cite{Gao16_jars} & 1542 & 77 & 97.53 & 79.55 & 80.83\\
GaborPCANet \cite{Gao16_grsl} & 2387 & \textbf{42} & 96.29 & 64.85 & 73.95\\
RPNet \cite{Xu18} & 795 & 258 & 98.39 & 85.11 & 86.35\\
LR-CNN \cite{Liu19} & 553 & 383 & 98.57 & 86.14 & 87.22\\
DCNet \cite{gao19jstars} & 447 & 392 & 98.72 & 87.40 & 89.04\\
Proposed DPDNet & \textbf{295} & 428 & \textbf{98.90} & \textbf{88.86} &\textbf{89.44}\\
\bottomrule
\end{tabular}
}
\label{table_Havana}
\end{table}

\textbf{Results on the Havana dataset.} Fig. \ref{result_Havana} shows the change maps on the Havana dataset. The corresponding evaluation metrics are listed in Table \ref{table_Havana}. As we can see, PCAKM, NR-ELM and GaborPCANet do not perform well since many unchanged regions are falsely classified into changed regions with high FP values. Moreover, the deep learning-based methods (LR-CNN, RPNet, and DCNet) can suppress the speckle noise to some extent by integrating spatial information. Therefore, the PCC, KC, and F1 values of these deep learning-based models have been greatly improved. However, some changed pixels are missed, and thus the FN values are relatively high. The proposed DPDNet obtains the best PCC, KC, and F1 values compared with other methods. Generally speaking, the comparisons demonstrate the effectiveness of the proposed method on the Havana dataset.

\begin{figure}[]
\centering
\includegraphics [width=3.3in]{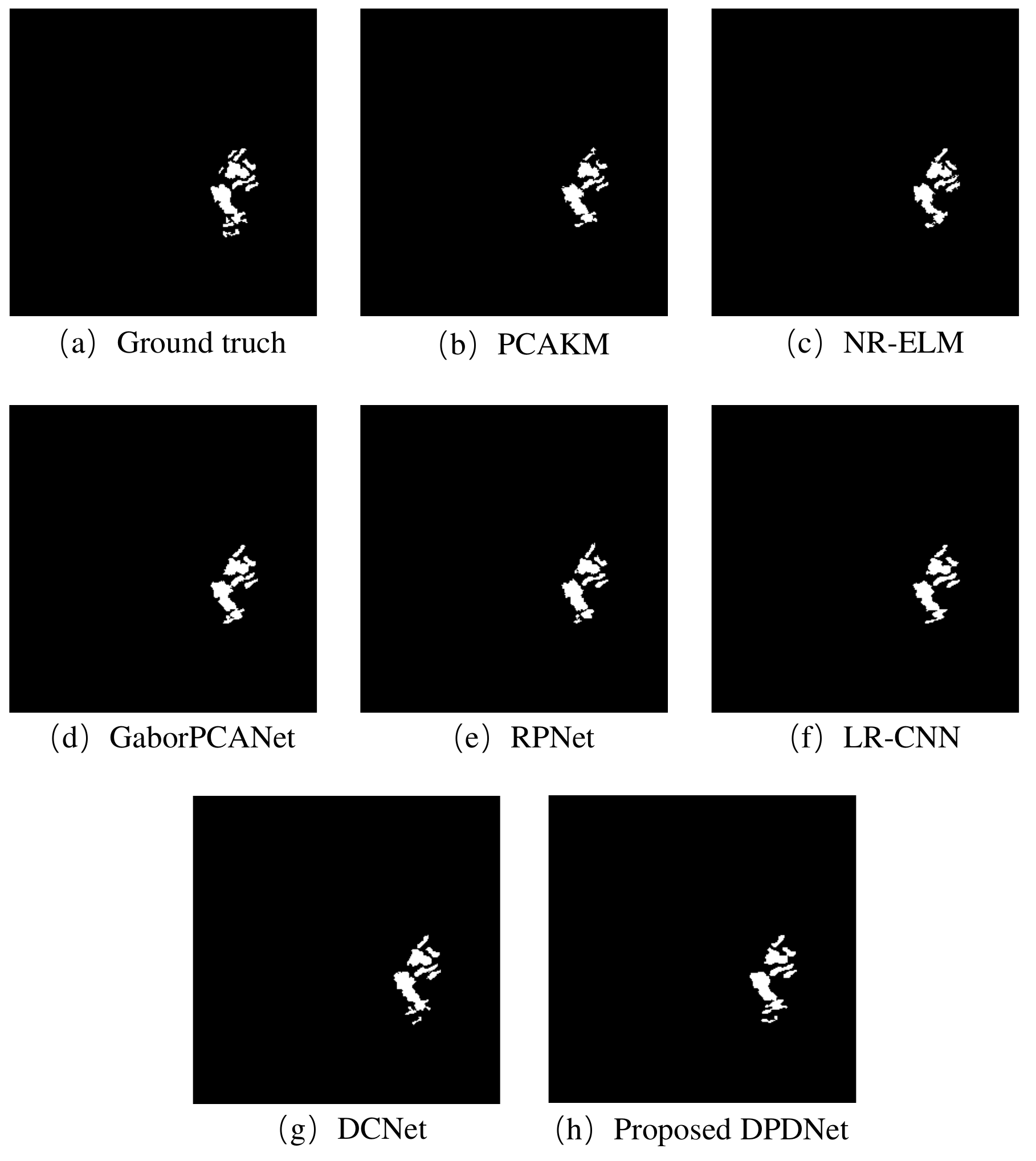}
\caption{Comparisons of the change maps generated by different methods on the Bern dataset. (a) Ground truth image. (b) Result by PCAKM \cite{Celik09}. (c) Result by NR-ELM \cite{Gao16_jars}. (d) Result by GaborPCANet \cite{Gao16_grsl}. (E) Result by RPNet \cite{Xu18}. (F) Result by LR-CNN \cite{Liu19}. (g) Result by DCNet \cite{gao19jstars}. (h) Result by the proposed DPDNet.}
\label{result_Bern}
\end{figure}

\begin{table}[]
\centering
\caption{Change detection results of different methods on the Bern dataset}
\renewcommand\arraystretch{1.5}
\setlength{\tabcolsep}{1.0mm}{
\begin{tabular}{c|ccccc}
\toprule
Methods & ~ FP ~ & ~ FN ~ & ~ PCC(\%) ~ & KC(\%) ~ & F1(\%)\\ \midrule
PCAKM \cite{Celik09} & \textbf{95} & 258 & 99.61 & 83.36 & 83.56\\
NR-ELM \cite{Gao16_jars} & 104 & 228 & 99.63 & 84.63 & 84.81\\
GaborPCANet \cite{Gao16_grsl} & 131 & 193 & 99.64 & 85.41 & 85.59\\
RPNet \cite{Xu18} & 103 & 226 & 99.64 & 84.77 & 84.96\\
LR-CNN \cite{Liu19} & 97 & 216 & 99.65 & 85.54 & 85.71\\
DCNet \cite{gao19jstars} & 146 & 161 & 99.66 & 86.45 & 86.62\\
Proposed DPDNet & 131 & \textbf{160} & \textbf{99.68} & \textbf{87.08} &\textbf{87.24}\\
\bottomrule
\end{tabular}
}
\label{table_bern}
\end{table}

\begin{figure*}[h]
\centering
\includegraphics [width=6.5in]{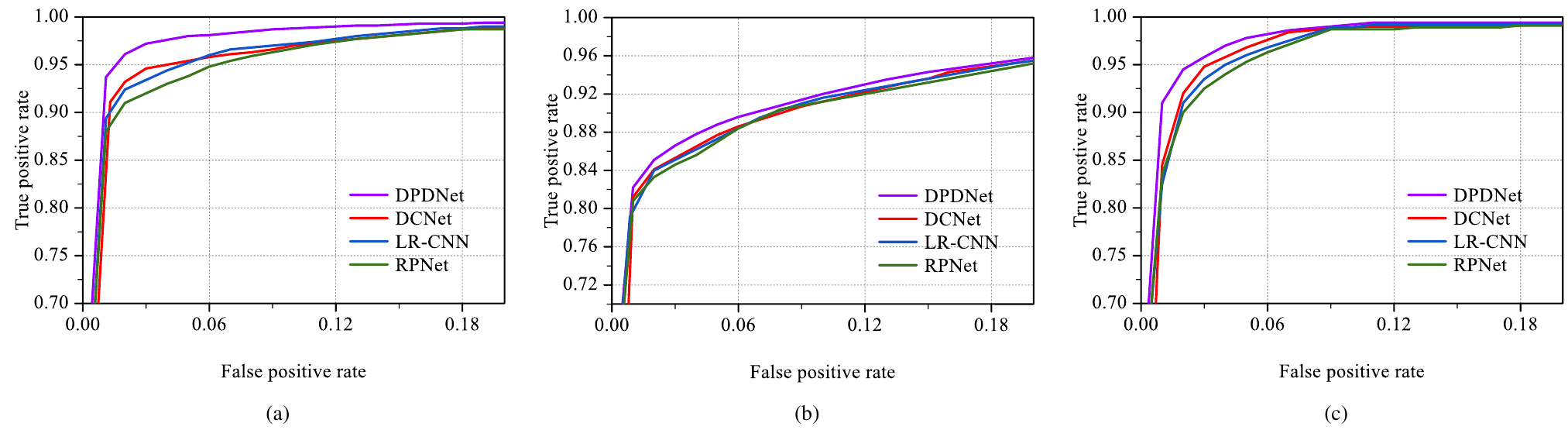}
\caption{ROC curves of the change detection results on different datasets. (a) ROC curves on the Florence dataset. (b) ROC curves on the Simulated dataset. (c) ROC curves on the Sulzberger dataset.}
\label{fig_roc}
\end{figure*}

\textbf{Results on the Bern dataset.} The visualized change detection results are shown in Fig. \ref{result_Bern}, and the evaluation metrics obtained by different methods are listed in Table \ref{table_bern}. It can be observed that the proposed DPDNet achieves the best performance in most metrics with the assistance of cleaned labels and multilevel features. Meanwhile, the result map of DPDNet is the most similar to the ground truth map. The satisfying performance on this open dataset proves the effectiveness of the proposed method in change detection task.

\subsection{ROC Curves of the Change Detection Results}

Fig. \ref{fig_roc} shows the receiver operating characteristic (ROC) curves of the change maps on the Florence, Simulated and Sulzberger datasets. It should be noted that the PCAKM, NR-ELM, GaborPCANet do not generate results with probability. Therefore, we only provide ROC curves of DPDNet, DCNet, LR-CNN, and RPNet.

The ROC curve of one good change detection method should be close to the top-left corner. On the three datasets, the ROC curves of the proposed DPDNet perform better than the other methods. The ROC curves of RPNet are closest to the bottom-right corner and have the worst performance. It is evident that the proposed DPDNet effectively improves the change detection performance by using the distinctive patch convolution. The corresponding area under the ROC curve (AUC) is provided in Table \ref{table_auc}. The higher the AUC, the better the performance of the model at distinguishing between the changed and unchanged pixels. We can observe that the DPDNet obtains the best AUC, which also demonstrates the effectiveness of the proposed method.

\begin{table}[htb]
\centering
\caption{AUCs of different methods on the Florence, Simulated and Sulzberger datasets}
\renewcommand\arraystretch{1.5}
\begin{tabular}{c|ccc}
\toprule
Methods & Florence & Simulated & Sulzberger\\
\midrule
RPNet \cite{Xu18} & 0.9831 & 0.9585 & 0.9867 \\
LR-CNN \cite{Liu19} & 0.9847 & 0.9603 & 0.9878\\
DCNet \cite{gao19jstars} & 0.9852 & 0.9602 & 0.9885 \\
Proposed DPDNet & 0.9910 & 0.9632 & 0.9912 \\
\bottomrule
\end{tabular}
\label{table_auc}
\end{table}

\begin{table}[htb]
\centering
\caption{Comparison among different change detection methods of the evaluation time (in seconds)}
\renewcommand\arraystretch{1.5}
\begin{tabular}{c|ccccc}
\toprule
Methods & Florence & Simulated & Sulzberger \\ \midrule
PCAKM \cite{Celik09} & 3.52 & 3.41 & 3.26 \\
NR-ELM \cite{Gao16_jars} & 24.62 & 23.81 & 22.75\\
GaborPCANet \cite{Gao16_grsl} & 441.68 & 435.80 & 431.76 \\
RPNet \cite{Xu18} & 158.62 & 155.78 & 153.30 \\
LR-CNN \cite{Liu19} & 482.53 & 478.33 & 465.90 \\
DCNet \cite{gao19jstars} & 492.01 & 488.62 & 481.80 \\
Proposed DPDNet & 165.15 & 163.24 & 162.87 \\
\bottomrule
\end{tabular}
\label{time}
\end{table}

\subsection{Analysis of the Computation Time Requirement}

Table \ref{time} shows the comparison among different change detection methods of the evaluation time on the Florence, Simulated and Sulzberger datasets. The traditional methods such as PCAKM and NR-ELM, cost less time, due to its relatively simple workflow. However, their performances are not as good as deep learning-based models. By contrast, the time consumptions of GaborPCANet, LR-CNN and DCNet are about more than twice higher than the proposed DPDNet. This demonstrates the computational efficiency of the DPDNet. The DPDNet requires much less computation time owing to the prefixed convolution kernels. Compared with RPNet, DPDNet costs more time to clean the label noise, while this further improves the experimental results. All experiments are carried out on a desktop PC with an Intel Xeon E5-2620 processor and an NVIDIA GTX 1080Ti GPU.

\section{Conclusions}

In this paper, we present a novel SAR image change detection model called Dual Path Denoising Network (DPDNet), which can suppress the speckle noise and label noise simultaneously. The DPDNet is comprised of two branches. One branch cleans the label noise via random label propagation, and the other branch combines shallow and deep features by using distincitve path convolution. In particular, the attention mechanism is used to select distinctive pixels in the feature map. Patches around these pixels are selected as convolution kernels. It does not require many training samples for parameter optimization, and therefore the proposed DPDNet is of high computational efficiency. The experimental results on five multitemporal SAR datasets demonstrate that the DPDNet can achieve substantially higher accuracy over state-of-the-art methods.

\begin{IEEEbiography}[{\includegraphics[width=1in,height=1.25in,clip,keepaspectratio]{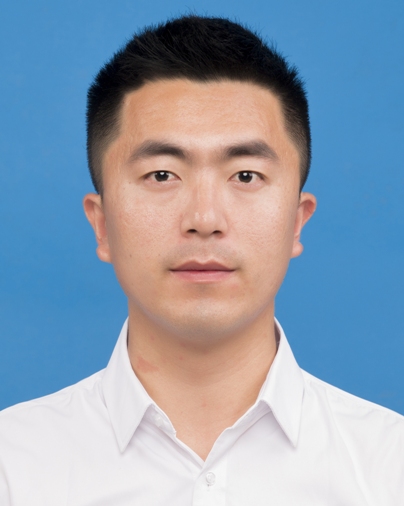}}]{Junjie Wang}
received the B.Sc. degree in computer science from Ocean University of China, Qingdao, China, in 2018. He is currently pursuing the M.Sc. degree in computer science and applied remote sensing with the School of Information Science and Technology, Ocean University of China, Qingdao, China.

His current research interests include computer vision and remote sensing image processing.

\end{IEEEbiography}

\begin{IEEEbiography}[{\includegraphics[width=1in,height=1.25in,clip,keepaspectratio]{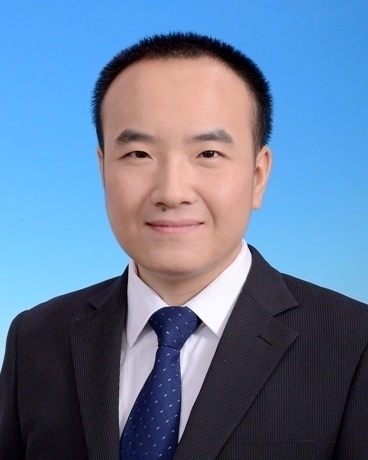}}]{Feng Gao} (Member, IEEE)
received the B.Sc degree in software engineering from Chongqing University, Chongqing, China, in 2008, and the Ph.D. degree in computer science and technology from Beihang University, Beijing, China, in 2015.

He is currently an Associate Professor with the School of Information Science and Engineering, Ocean University of China. His research interests include remote sensing image analysis, pattern recognition and machine learning.

\end{IEEEbiography}

\begin{IEEEbiography}[{\includegraphics[width=1in,height=1.25in,clip,keepaspectratio]{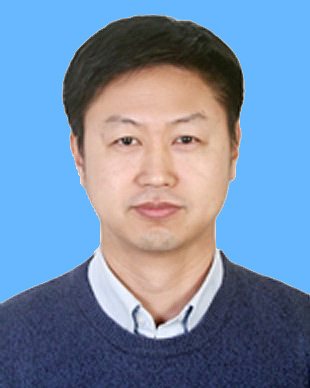}}]{Junyu Dong}
 (Member, IEEE) received the B.Sc. and M.Sc. degrees from the Department of Applied Mathematics, Ocean University of China, Qingdao, China, in 1993 and 1999, respectively, and the Ph.D. degree in image processing from the Department of Computer Science, Heriot-Watt University, Edinburgh, United Kingdom, in 2003.

He is currently a Professor and Dean with the School of Computer Science and Technology, Ocean University of China. His research interests include visual information analysis and understanding, machine learning and underwater image processing.
\end{IEEEbiography}

\begin{IEEEbiography}[{\includegraphics[width=1in,height=1.25in,clip,keepaspectratio]{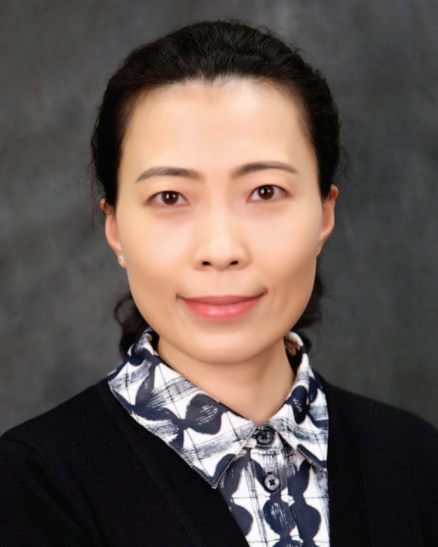}}]{Qian Du}
(Fellow, IEEE) received the Ph.D. degree in electrical engineering from the University of Maryland at Baltimore, Baltimore, MD, USA, in 2000.

She is currently the Bobby Shackouls Professor with the Department of Electrical and Computer Engineering, Mississippi State University, Starkville, MS, USA. Her research interests include hyperspectral remote-sensing image analysis and applications, and machine learning.

Dr. Du was the recipient of the 2010 Best Reviewer Award from the IEEE Geoscience and Remote Sensing Society (GRSS). She was a Co-Chair for the Data Fusion Technical Committee of the IEEE GRSS from 2009 to 2013, the Chair for the Remote Sensing and Mapping Technical Committee of International Association for Pattern Recognition from 2010 to 2014, and the General Chair for the Fourth IEEE GRSS Workshop on Hyperspectral Image and Signal Processing: Evolution in Remote Sensing held at Shanghai, China, in 2012. She was an Associate Editor
for the \textsc{IEEE Journal of Selected Topics in Applied Earth Observation and Remote Sensing}, \emph{Journal of Applied Remote Sensing}, and \textsc{IEEE Signal Processing Letters}. From
2016 to 2020, she was the Editor-in-Chief of the \textsc{IEEE Journal of Selected Topics in Applied Earth Observation and Remote Sensing}. She is currently a member of the IEEE Periodicals Review and Advisory Committee and SPIE Publications Committee. She is a Fellow of SPIE-International Society for Optics and Photonics (SPIE).

\end{IEEEbiography}

\begin{IEEEbiography}[{\includegraphics[width=1in,height=1.25in,clip,keepaspectratio]{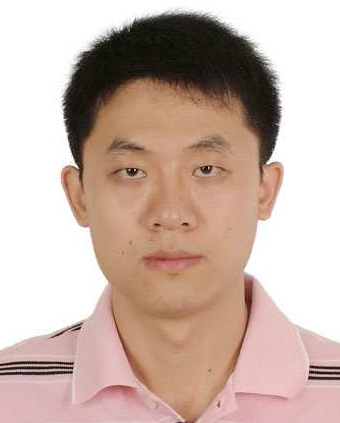}}]{Heng-Chao Li}
(Senior Member, IEEE) received the B.Sc. and M.Sc. degrees from Southwest Jiaotong University, Chengdu, China, in 2001 and 2004, respectively, and the Ph.D. degree from the Graduate University of Chinese Academy of Sciences, Beijing, China, in 2008, all in information and communication engineering. He is currently a Full Professor with the School of Information Science and Technology, Southwest Jiaotong University. His research interests include statistical analysis of synthetic aperture radar (SAR) images, remote sensing image processing, and pattern recognition.

Dr. Li is an Editorial Board Member of the \emph{Journal of Southwest Jiaotong University} and the \emph{Journal of Radars}. He was a recipient of the 2018 Best Reviewer Award from the IEEE Geoscience and Remote Sensing Society for his service to the \textsc{IEEE Journal of Selected Topics in Applied Earth Observations and Remote Sensing} (JSTARS). Moreover, he has served as a Guest Editor for the Special Issues of the \emph{Journal of Real-Time Image Processing}, the IEEE JSTARS, and the IEEE JMASS, a Program Committee Member for the 26th International Joint Conference on Artificial Intelligence (IJCAI-2017), and the 10th International Workshop on the Analysis of Multitemporal Remote Sensing Images (MULTITEMP-2019), and the Session Chair for the 2017 International Geoscience and Remote Sensing Symposium (IGARSS-2017), and the 2019 Asia-Pacific Conference on Synthetic Aperture Radar (APSAR-2019). He is serving as an Associate Editor for the IEEE JSTARS.
\end{IEEEbiography}

\end{document}